\documentclass{aa}
\usepackage{graphics}
\usepackage{latexsym}
\usepackage{astron}
\usepackage{amssymb}
\usepackage{amsmath}
\begin{document} 
 
   \title{Chemical abundances in 43 metal-poor stars
   \thanks{Based on observations carried out at the European Southern
           Observatory, La Silla, Chile},
   \thanks{Tables 1 and 3 are only, and Tables 2 and 4 are also available in
           electronic form at the CDS via anonymous ftp to cdsarc.u-strasbg.fr
           (130.79.128.5) or via http://cdsweb.u-strasbg.fr/Abstract.html }
         }

   \author{Karin Jonsell\inst{1}
   \and    Bengt Edvardsson\inst{1}
   \and    Bengt Gustafsson\inst{1}
   \and    Pierre Magain\inst{2}
   \and    Poul Erik Nissen\inst{3}
   \and    Martin Asplund\inst{4}}
   \offprints{K. Jonsell}

   \institute{Department of Astronomy and Space Physics, Uppsala Astronomical
              Observatory, Box 515, S-751 20 Uppsala, Sweden
              (firstname.lastname@astro.uu.se)
   \and       Institut d'Astrophysique et de G\'eophysique, Universit\'e\
              de Li\`ege, All\'ee du 6 Ao\^ut 17, B-4000 Li\`ege, Belgium
              (Pierre.Magain@ulg.ac.be)
   \and       Institute of Physics and Astronomy, University of Aarhus,
              DK-8000 Aarhus C, Denmark (pen@phys.au.dk)
   \and       Research School of Astronomy and Astrophysics, Mount Stromlo 
              Observatory, Cotter Road, Weston, ACT 2611, Australia 
              (martin@mso.anu.edu.au)}

   \date{Received 31 January 2005  / Accepted 26 April 2005}

   \authorrunning{K. Jonsell et al.}

   \abstract{
We have derived abundances of O, Na, Mg, Al, Si, Ca, Sc, Ti, V, Cr, Fe, Ni
and Ba for 43 metal-poor field stars in the solar neighbourhood, most of 
them subgiants or turn-off-point stars, with iron abundances [Fe/H] ranging from 
$-0.4$ to $-3.0$. About half of this sample has not been 
analysed spectroscopically in detail before. Effective
temperatures were estimated from $uvby$ photometry, and surface gravities 
primarily from Hipparcos parallaxes. The analysis is differential relative
to the Sun, and carried out with plane-parallel MARCS models.
Various sources of error are discussed
and found to contribute a total error of about 0.1--0.2\,dex for most elements,
while relative abundances, such as [Ca/Fe], are most probably more accurate.
For the oxygen abundances, determined in an NLTE analysis of the 7774\,\AA\ 
triplet lines, the errors may be somewhat larger. 
We have made a detailed comparison with other similar studies and traced the
reasons for the, in most cases, relatively small differences. 

Among the results we note the following:
We find [O/Fe] to possibly increase beyond [Fe/H]\,$=-1.0$, though
considerably less than obtained by others from abundances based on OH lines.
We do not trace any tendency for strong overionization of iron.
We find the excesses, relative to Fe and the Sun, of the $\alpha$ elements Mg, 
Si and Ca to be smaller than those of O. We discuss some indications that also
the abundances of $\alpha$ elements relative to Fe vary and
the possibility that some of the scatter around the trends in abundances relative
to iron may be real.
This may support the idea that the formation of the Halo stars occurred in
smaller systems with different star formation rates. We verify the finding by
Gratton et al. (2003b) that stars that do not participate in the rotation
of the galactic disk show a lower mean and larger spread in [$\alpha$/Fe]
than stars participating in the general rotation.
The latter stars also seem to show some correlation between [$\alpha$/Fe] and
rotation speed. We trace some stars with peculiar abundances, among those two Ba
stars, \object{HD\,17072} and \object{HD\,196944},
the latter already known to be rich in $s$ elements. We finally advocate that a
spectroscopic study of a larger sample of halo stars with well-defined 
selection criteria is very important, in addition to the very considerable
efforts that various groups have already made.

 }

     \maketitle


\section{Introduction}
One important method of studying the formation and early evolution of galaxies
is to explore the properties of the remains of these early events in the
Milky Way system.
Since the pioneering work by Eggen, Lynden-Bell \& Sandage (1962),
investigations of stellar abundances and stellar dynamics have
been combined in a great number of studies in attempts to reconstruct the
formation history of the stellar populations. While the homogeneous collapse
scenario of Eggen et al. predicted metallicity and age gradients related to
the kinematics, the globular clusters were found to show a considerable
spread in metal abundances, independent of their distance from the Galactic
centre (Searle \& Zinn, 1978). This instead suggested a scenario of galaxy
formation where individual protogalactic systems evolved independently before
merging into a larger system. There are a number of indications that the
infall and accretion of material after the first collapse phase has played
an important role in our Galaxy: the existence of high-latitude 
and high-velocity interstellar clouds, the relatively small proportion of
metal-poor solar-type stars (``the G-dwarf problem"), the large
scatter in the age-metallicity relation (if this relation exists at all) for
the Galactic disk, and the finding of the presently merging Sagittarius dwarf. 

More recently, important clues have come from abundance differences 
in different stellar populations.
Thus, Gratton et al. (1996, 2000) and Fuhrmann (1998) have traced significant
[$\alpha$/Fe] ratio (i.e., e.g. Mg, Si, and Ca vs. Fe)
differences between the Thick and the Thin Galactic disks, suggesting a time
difference between the two systems and a period of little or no star-formation
in between, possibly marking different large-scale infall events.
Different kinematic properties at a given [Fe/H],
may be traced in the disk (see also Bensby et al. 2003).
Also for the halo stars, two different populations have been traced, as
reflected in kinematic properties and [$\alpha$/Fe] ratios
(Nissen \& Schuster 1997, Gratton et al. 2003a, 2003b).
Gratton et al. distinguish between one population with positive velocity 
in orbits around the galactic centre and with close correlations 
between rotational velocity, metallicity and [$\alpha$/Fe], and a second 
population with small or retrograde orbital velocity with larger scatter in the
latter respects.
The star-to-star scatter in relative abundances, such as in
[$\alpha$/Fe] at a given [Fe/H] for halo stars may also be used to obtain
information on the supernovae, star-formation processes and gas dynamics in the
early Galaxy, since supernovae with different initial masses produce different
relative amounts of the heavy elements (see Nissen et al. 1994,
Karlsson \& Gustafsson 2001, Carretta et al. 2002, Arnone et al. 2004,
Cayrel et al. 2004).

In order to study the star-to-star scatter, as well as the possible existence 
of fine structure in the halo-population and the transition to the thick disk
population in terms of abundances and kinematics, the present project was 
initiated with observations more than a decade ago.
About half of our programme stars have more recently been independently
studied.
Our results give checks or support of results obtained by others
and add some knowledge of significance for the understanding of Halo formation.

\section{Observations and data reductions}
\label{Sobservations}
43 stars were chosen, mainly from the Olsen (1983) catalogue
of Str\"{o}mgren $uvby-\beta$ photometry and the Olsen (1993) catalogue of
G-dwarfs. The parameters for the stars lies within the following intervals:
$4\fm2 \leq V \leq 9\fm1$, $0\fm30 \leq (b-y) \leq 0\fm51$ and [Me/H]\,$<-0.4$.
The stars all have $-73^\circ \leq \delta \leq +21^\circ$.

The observations were performed in 1987 and 1988
at the ESO 1.4\,m Coud\'{e} Auxiliary Telescope,
CAT, with the short camera of the Coud\'{e} Echelle Spectrometer, CES, equipped
with the RCA CCD no.\,9. The setup made it possible to obtain a spectral
resolving power of 60,000 and a signal to noise ratio over about 200.
To get a reasonably large sample of unblended lines of important elements four
different wavelength regions were observed: 5670--5720, 6120--6185, 
7750--7820 and 8710--8780\,\AA.

The CCD used in the programme, RCA no.\,9, was not fully linear at the time of
observation, as found by one of us in a careful analysis including simulations
of the effects on measured equivalent widths (Gosset \& Magain 1993).
To correct for this nonlinearity our equivalent widths, $W$,
were reduced:
\begin{equation}
W_{\rm corrected} = 0.96 \cdot W_{\rm measured}.
\label{Eeqwcorr}
\end{equation}

Strong interference fringes appeared in the infrared region.
Dome and hot-star flat-fields were used to correct for these.

\begin{figure}
\centering
\resizebox{\hsize}{!}{\includegraphics*{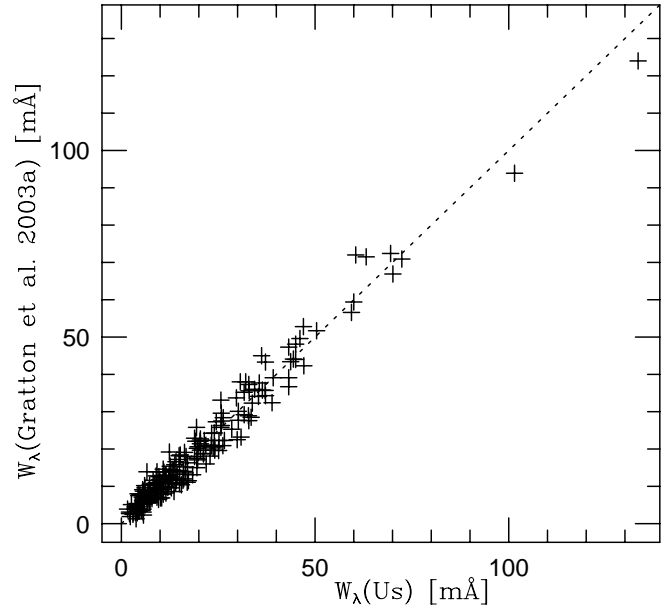}}
\caption[]{\label{Fcompeqw}
Comparison of equivalent width measurements with those of
Gratton et al. (2003a).
The dotted line indicates the 1-to-1 relation.
}
\end{figure}

Up to 62 usable absorption lines were measured. For infrared lines with
equivalent widths under 40\,m{\AA}, the widths were measured with a Gaussian fit.
Lines stronger than this were fitted with a Voigt profile. Lines in the visual
region of the stellar spectra were measured by a weighted mean between a
Gaussian fit and a pure numerical integration.
The weighting was estimated by eye
with respect to suspected blends, line strength etc.
The widths for \object{HD\,140283} and \object{HD\,196944} in the 7780\,{\AA}
region are the averages for two spectra. 
We compare in Fig.\,\ref{Fcompeqw} our equivalent widths for 212 lines with
measurements by Gratton et al. (2003a).
The mean difference (Us-Gratton et al.) is $+0.1$\,m\AA\ with a standard
deviation of 3.2\,m\AA.
The adopted equivalent widths are given in Table\,1
(only electronically available).

\section{Analysis}
\subsection{Model atmospheres}
\subsubsection{Properties of the models}
\label{model}
A version of the Uppsala model atmosphere code MARCS was used to calculate
the atmospheres for the programme stars, see Asplund et al. (1997).

\subsubsection{Fundamental parameters of the atmospheres}
\label{Sfundamentalparameters}
The basic stellar data are presented in Table\,\ref{Tstardata}.
\setcounter{table}{1}
\begin{table*}
\begin{scriptsize}
\caption[]{\label{Tstardata}
Data for the 43 stars in the project.
Column 1: Stellar identification.
Column 2: Visual apparent magnitude.
Column 3: Colour index $(b-y)$.
Column 4: Balmer discontinuity index $c_1$. 
Column 5: References for the photometry: 
1) Olsen (1983), 
2) Schuster \& Nissen (1988), 
3) Lindgren H. (unpublished), 
4) Olsen (1993), 
5) Hauck \& Mermilliod (1998),
6) Gr\o nbech \& Olsen (1976, 1977), 
7) SIMBAD.
Column 6, 7 and 8: Stellar space velocities relative to the LSR,
see Sect.\,\ref{Sspacevelocities}. 
Column 9: Stellar masses in units of the Solar mass,
see Sect.\,\ref{Sfundamentalparameters}. 
Column 10: The effective temperatures.
except for \object{HD\,196944}.
Column 11: The metallicities determined from spectroscopy,
see Sect.\,\ref{Sfundamentalparameters}. 
Column 12: Gravitational parameters from Hipparcos parallaxes and
evolutionary tracks (VandenBerg et al. 2000), except for \object{HD\,196944}.
Column 13: The classification of the stars according to the SIMBAD database. 
Column 14: Ba = barium star, b = binary (Nordstr\"{o}m et al. 2004),
b? = possible binary (Nissen et al. 1997),
v = variable (Cayrel de Strobel et al. 1997). }
\begin{flushleft} 
\begin{tabular*}{18.0cm}{@{\extracolsep\fill}l|cccl|rrr|llll|ll}
HD & $V$ &$(b-y)$& $c_1$ & Ref. & $U$  & $V$  & $W$  & Mass      & $T_{\rm eff}$ & [Fe/H] & $\log g$ & Classification & Notes\\ 
   & mag & mag   & mag   &      & km/s & km/s & km/s & $M_\odot$ & K             &  dex   &          &                &      \\
\hline
 17072 & 6.594 & 0.441 & 0.453 & 1,2   &    5 &  $-$66 &  $-$16 & 0.80$^1$ & 5428 & $-$0.98 & 2.65 & G2wF5    & Ba       \\ 
 25704 & 8.124 & 0.371 & 0.274 & 1,2   & $-$119 &  $-$62 &   $-$1 & 0.75 & 5758 & $-$0.97 & 4.16 & F7V       &          \\ 
 49301 & 8.112 & 0.371 & 0.408 & 1     &   27 &  $-$47 &   21 & 0.80$^1$ & 5832 & $-$0.82 & 3.11 & F5V       &          \\ 
 59374 & 8.478 & 0.363 & 0.273 & 2     &  $-$49 & $-$120 &    0 & 0.74 & 5811 & $-$0.93 & 4.36 & F8V       &          \\ 
 61902 & 8.232 & 0.327 & 0.357 & 1,2,3 &   77 &  $-$16 &  $-$38 & 0.99 & 6105 & $-$0.71 & 4.04 & F5/6wF2  & v       \\ 
 63598 & 7.947 & 0.358 & 0.267 & 2     &   27 & $-$103 &   10 & 0.86 & 5845 & $-$0.84 & 4.24 & G2V       & b    \\ 
 76932 & 5.806 & 0.355 & 0.294 & 1,2,3 &  $-$38 &  $-$85 &   77 & 0.87 & 5875 & $-$0.89 & 4.13 & F7/F8IV/V & b    \\ 
 78747 & 7.717 & 0.385 & 0.284 & 1,3   &   20 &    7 &  $-$19 & 0.76 & 5684 & $-$0.80 & 4.22 & G0/G1V    &          \\ 
 79601 & 8.016 & 0.382 & 0.304 & 4     &  $-$10 &  $-$35 &  $-$29 & 0.80 & 5716 & $-$0.74 & 4.12 & G2V       &          \\ 
 80863 & 8.685 & 0.365 & 0.304 & 4     &  $-$37 &  $-$68 &  $-$22 & 0.80 & 5834 & $-$0.61 & 4.26 & F7/F8V    & b    \\ 
 84937 & 8.335 & 0.301 & 0.359 & 1,2,3 &  235 & $-$234 &    0 & 0.75 & 6310 & $-$1.96 & 4.04 & sdF5      &      \\ 
 91121 & 8.780 & 0.390 & 0.352 & 5,7   &   75 & $-$176 &  $-$77 & 0.70 & 5673 & $-$1.08 & 3.88 & G1WF2/5 V &          \\ 
 94028 & 8.227 & 0.342 & 0.254 & 1,2,3 &  $-$26 & $-$135 &   18 & 0.70 & 5934 & $-$1.39 & 4.23 & F4V       &      \\ 
 97320 & 8.170 & 0.338 & 0.301 & 1,3   &   84 &  $-$18 &  $-$31 & 0.72 & 5983 & $-$1.23 & 4.17 & F3V       &          \\ 
 99682 & 8.645 & 0.342 & 0.373 & 4     &  $-$65 &  $-$15 &   $-$8 & 0.97 & 6034 & $-$0.45 & 4.06 & F6V       &          \\ 
101976 & 8.310 & 0.413 & 0.315 & 4     &    1 &   27 &   66 & 0.95 & 5554 & $-$0.49 & 3.87 & G1V       &          \\ 
102200 & 8.740 & 0.330 & 0.300 & 2,3   &  100 & $-$136 &   16 & 0.76 & 6037 & $-$1.23 & 4.13 & F2V       &          \\ 
108317 & 8.036 & 0.447 & 0.291 & 4     & $-$127 & $-$106 &  $-$11 & 0.80$^1$ & 5332 & $-$2.19 & 2.75 & G0        & b    \\ 
111777 & 8.490 & 0.395 & 0.257 & 2     & $-$109 &  $-$82 &  $-$39 & 0.71 & 5606 & $-$0.85 & 4.36 & G1/G2     &          \\ 
116064 & 8.807 & 0.345 & 0.273 & 2,3   &  $-$94 & $-$222 &  120 & 0.70$^2$ & 5945 & $-$1.88 & 4.28 & sdF0      &  \\ 
120559 & 7.970 & 0.423 & 0.203 & 1,2   &  $-$18 &  $-$43 &  $-$29 & 0.70 & 5406 & $-$0.99 & 4.59 & G5WF8 V   & b    \\ 
122196 & 8.736 & 0.349 & 0.331 & 1,3   & $-$164 & $-$136 &   20 & 0.84 & 5934 & $-$1.69 & 3.92 & sdF5      &          \\ 
124785 & 8.666 & 0.388 & 0.328 & 1,3   &  $-$29 & $-$105 &   46 & 0.86 & 5694 & $-$0.69 & 3.86 & F8/G0WF5 &          \\ 
126793 & 8.208 & 0.367 & 0.290 & 1,3   &   $-$4 &    6 &   $-$7 & 0.80 & 5800 & $-$0.81 & 4.18 & G0WF3/5  &          \\ 
128279 & 8.034 & 0.465 & 0.268 & 1,3   &   23 &  $-$92 & $-$265 & 0.80$^1$ & 5216 & $-$2.16 & 2.94 & G0        &          \\ 
132475 & 8.558 & 0.393 & 0.283 & 1,2,3 &   52 & $-$361 &   61 & 0.70$^2$ & 5619 & $-$1.53 & 3.75 & F5/F6V  & v       \\ 
140283 & 7.213 & 0.380 & 0.287 & 1,2   & $-$238 & $-$248 &   50 & 0.77 & 5751 & $-$2.26 & 3.71 & sdF3      & v        \\ 
142945 & 8.011 & 0.389 & 0.251 & 1     &   70 &  $-$41 &    4 & 0.70 & 5633 & $-$1.00 & 4.35 & F8V       & b   \\ 
144450 & 8.084 & 0.485 & 0.245 & 4     &   $-$5 &  $-$76 &   22 & 0.70$^3$ & 5077 & $-$1.17 & 3.42 & G6V       &          \\ 
145417 & 7.520 & 0.510 & 0.170 & 1,2   &  $-$40 &  $-$85 &  $-$22 & 0.60$^2$ & 4908 & $-$1.36 & 4.69 & G8/K0 V(W) &          \\ 
160617 & 8.733 & 0.345 & 0.335 & 2,3   &   65 & $-$210 &  $-$87 & 0.77 & 5967 & $-$1.77 & 3.79 & Fw        &          \\ 
166913 & 8.221 & 0.327 & 0.302 & 1,3   &  $-$41 &  $-$40 &   77 & 0.72 & 6065 & $-$1.54 & 4.13 & F6:Vw     &          \\ 
188510 & 8.834 & 0.416 & 0.163 & 2     & $-$142 & $-$108 &   70 & 0.60$^2$ & 5423 & $-$1.62 & 4.47 & G5Vw      & b   \\ 
193901 & 8.659 & 0.382 & 0.219 & 1,2   & $-$146 & $-$240 &  $-$67 & 0.70$^2$ & 5657 & $-$1.16 & 4.46 & F7V       &          \\ 
194598 & 8.354 & 0.344 & 0.269 & 2     &  $-$66 & $-$271 &  $-$24 & 0.70 & 5928 & $-$1.14 & 4.23 & F7V$-$VI    &          \\ 
196892 & 8.245 & 0.349 & 0.304 & 1,2,3 &    8 & $-$125 &  $-$24 & 0.78 & 5912 & $-$1.09 & 4.11 & F6V       & b?       \\ 
196944 & 8.398 & 0.449 & 0.348 & 1     & $-$142 & $-$131 &  $-$15 & 0.80$^1$ & 5353$^4$ & $-$2.23 & 1.70$^4$ & G2-5II & Ba, b \\ 
199289 & 8.293 & 0.363 & 0.261 & 1,2   &  $-$31 &  $-$62 &  $-$18 & 0.70$^2$ & 5800 & $-$1.10 & 4.21 & F5V       & b?       \\ 
200654 & 9.097 & 0.460 & 0.271 & 2     & $-$265 & $-$384 & $-$138 & 0.80$^{1,3}$ & 5340 & $-$2.99 & 2.96 & G:w       &          \\ 
201891 & 7.379 & 0.359 & 0.254 & 1,2,3 &  102 & $-$110 &  $-$52 & 0.70 & 5821 & $-$1.14 & 4.20 & F8V$-$VI    & b?       \\ 
203608 & 4.229 & 0.330 & 0.317 & 6     &   $-$3 &   49 &   13 & 0.88 & 6063 & $-$0.74 & 4.29 & F6V       & v       \\ 
211998 & 5.281 & 0.451 & 0.239 & 1,2   & $-$160 & $-$144 &  $-$68 & 0.89 & 5255 & $-$1.49 & 3.44 & A3V?+F9V  & b        \\ 
213467 & 8.531 & 0.484 & 0.288 & 1     & $-$223 & $-$220 &  $-$79 & 0.80$^{1,3}$ & 5099 & $-$1.43 & 2.92 & G5VWF3  &          \\ 
\hline
\end{tabular*}
\end{flushleft}
$^1$ Adopted mass 0.80\,$M_\odot$.
$^2$ The star is located on the lower side of the track.
$^3$ The star is located where the tracks are very compact.
$^4$ $T_{\rm eff}$, $\log g$ and $\xi_{\rm t}$ adopted from Za\u{c}s et al. (1998)
\end{scriptsize}
\end{table*}

{\it Effective temperatures.}
The effective temperatures of the programme stars were
derived from Str\"{o}mgren $uvby-\beta$ photometry collected from
sources given in the Table caption.
The empirical temperature calibration of Alonso et al. (1996) (their Eq.\,9)
was used to estimate the effective temperatures.
This calibration is based on the InfraRed Flux Method, a grid of Kurucz'
atmosphere models and a large sample of dwarfs and subdwarfs in the metallicity
range $-3\leq$\,[Fe/H]\,$\leq0.5$. 
In applying the calibration we iterated until consistency was achieved between
the calculated temperatures, the metallicities, the surface gravities and the
derived iron abundances of the stars, see also below.
The effective temperature of the star \object{HD\,196944} was adopted from
Za\u{c}s et al. (1998).
The reddening has been calculated for all stars from H$_\beta$, and was
found to be small for all stars but one, and reddening has
therefore been neglected in the analysis (see Sect.\,\ref{Stemperror}, below).

{\it Metallicities.}
The overall metallicities, [Me/H], of the stars were first approximated
by using a calibration of Str\"{o}mgren photometry made by
Schuster \& Nissen.
The metallicities were changed until consistency was
achieved with the derived iron abundances, [Fe/H], from lines of Fe\,{\sc i},
the surface gravities and the effective temperatures.

{\it Surface gravities.}
The surface gravities were calculated from Hipparcos parallaxes 
in the same way as described by Nissen et al. (1997), their Eq.\,(3).
An age of 14\,Gyr was assumed for all stars.
The stellar masses were estimated from the evolutionary tracks of
VandenBerg et al. (2000).
For some stars, which were not faint enough for these tracks, 0.80\,$M_\odot$
was adopted.
Some stars were located on the lower side of the track, and some stars resided
in the crowded part of the giant branch.
In these cases, the masses will have higher uncertainties.
The visual magnitudes were adopted from different sources, the bolometric
corrections were determined from Alonso et al. (1995) and the parallaxes adopted
from the Hipparcos catalogue (ESA 1997). The temperatures, metallicities,
iron abundances and gravities were iterated until consistency was achieved.
The surface gravity of the star \object{HD\,196944} was set to $\log g=1.7$
according to Za\u{c}s et al. (1998).

{\it Microturbulence parameters.}
The microturbulence parameter, $\xi_{\rm t}$, represents the doppler broadening
of the lines by non-thermal small-scale motions in the stellar atmosphere.
Here the microturbulence parameter was set to $\xi_{\rm t}=1.5$\,km/s.
A microturbulence parameter of 1.9\,km/s was adopted from Za\u{c}s et al. (1998)
for the star \object{HD\,196944}.

\subsection{Stellar space velocities} 
\label{Sspacevelocities}
Galactic space velocities $U$, $V$ and $W$, in a right-handed system with $U$
directed towards the galactic center, were computed for the programme stars with
the transformations given by Johnson \& Soderblom (1987). Parallaxes and proper
motions were adopted from the Hipparcos catalogue (ESA 1997). Accurate
radial velocity data for 39 of our 43 stars were kindly supplied in advance of
publication by Birgitta Nordstr\"om, Copenhagen (Nordstr\"om et al. 2004).
For the remaining four stars the radial velocities were obtained from the CDS
database (Egret 1986).
The resulting velocities relative to the LSR are given in Table\,\ref{Tstardata},
where we have used the solar velocities relative to the LSR $U_\odot=+10.00$,
$V_\odot=+5.25$ and $W_\odot=+7.17$\,km/s of Dehnen \& Binney (1998).  
\begin{figure}
\centering
\resizebox{\hsize}{!}{\includegraphics*{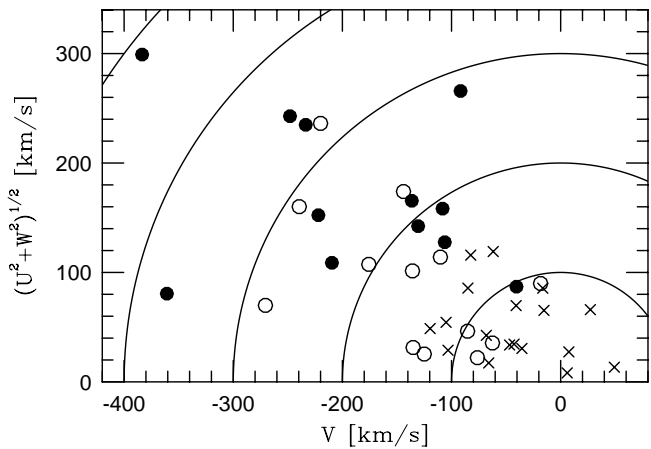}}
\caption[]{\label{Fvelocities}
A Toomre diagram for the stars, where the symbols have been coded
according to the overall stellar metallicity:
$-1.0\leq$\,[Fe/H] (crosses), $-1.5\leq$\,[Fe/H]\,$<-1.0$ (open circles)
and [Fe/H]\,$<-1.5$ (filled circles). The curves connect loci with identical
total velocities relative the LSR, and the diagram shows an obvious
anticorrelation between overall metallicity and velocity relative to the LSR.}
\end{figure}

Fig.\,\ref{Fvelocities} shows the kinematic Toomre diagram for our sample.
The figure displays the well-known relation between overall metallicity and
velocity relative to the local standard of rest.

\subsection{Abundance analysis}
\subsubsection{The properties of the analysing program}
\label{Sanalysingprogram}
The Uppsala programme EQWIDTH was used to analyse the measured stellar 
absorption lines.
In this programme LTE is assumed but a source function is adopted
which properly allows for continuum scattering.
This is important mainly for UV spectra of metal-poor stars.
For a given model atmosphere and for an adequate number of wavelength points
EQWIDTH solves the equation of radiative transfer, integrates an 
equivalent width, compares with the observed width and then
determines the corresponding abundance by iteration.

The lines chosen were selected to be essentially free of blends. Since most
lines are weak both in the solar and stellar spectra, the effects on the
equivalent widths by isotope splitting (IS) and hyperfine-structure splitting
(hfs) should not be important.
The exception may be the Ba\,{\sc ii} line (at 6141.71\,\AA) which is blended
with a line of Fe\,{\sc i}, and which shows both hfs and IS.
For this line we made detailed synthetic-spectrum calculations with the
programme BSYN, a sister programme to EQWIDTH, taking the iron-line blend and
the different line components into consideration with hfs and IS data
from McWilliam (1998), assuming solar isotope ratios.
We found, however, that the IS and hfs splittings are negligible for this line.

The oxygen IR triplet lines, used here for determining the oxygen abundances,
are known to be severely affected by departures from LTE (e.g. Kiselman 1991,
1993, Nissen et al. 2002).  We have therefore solved the statistical-equilibrium 
equations for an oxygen model atom with 23 levels and 65 radiative transitions, 
kindly provided by Dan Kiselman.
For method and atomic data; see Nissen et al. (2002).
These calculations were made individually for
each star with appropriate model atmospheres.
The corrections in abundances
range from $-0.07$ to $-0.33$\,dex and seem only little affected by
the uncertainties in hydrogen-collision cross sections (cf. Nissen et al. 2002).

\subsubsection{Atomic line data}
\label{linedata}
The Vienna Atomic Line Data Base, VALD (Kupka et al. 1999), was used
for gathering atomic line data, such as wavelengths, $\lambda$,
excitation energies, $\chi_{\rm lower}$,
and radiation damping constants, $\gamma_{\rm rad}$.
For some lines of sodium, magnesium and silicon VALD had no radiation damping
constants listed, and they were then taken from the tables of Kurucz (1989).
For the barium line the radiation damping
constant was calculated from lifetimes of relevant energy levels identified
in the tables of Bashkin \& Stoner (1975).
The line data are given in Table\,3 (only available in electronic form).

\subsubsection{Pressure broadening}
\label{Spressurebroadening}
To deal with collisional broadening induced by neutral hydrogen, we have applied
the data of Barklem et al. (2000), based on quantum mechanical calculations for
46 of 62 lines. 
For lines not covered by Barklem et al. the Uns\"old (1955) approximation with 
enhancement factors was adopted. The enhancement factors applied in this 
investigation were adopted from the literature: Na\,{\sc i}: Holweger (1971), 
Si\,{\sc i}: Holweger (1973) and Fe\,{\sc ii}: Holweger et al. (1990). 
For Mg\,{\sc i} and Sc\,{\sc ii} a value of 2.50 was applied following 
M\"ackle et al. (1975).
The different broadening methods are compared in Sect.\,\ref{Slinebroadening}.

\subsubsection{Oscillator strengths}
\label{Soscillatorstrengths}
For determining the absolute abundance one needs the product of the 
statistical weight, $g$, and the oscillator strength, $f$, for the transition. 
We have performed a differential analysis relative to the Sun 
and determined astrophysical $\log gf$ values
by running the analysis programme EQWIDTH with known abundances, solar-flux
equivalent widths and a solar model atmosphere.
The photospheric solar abundances, $\log \epsilon_{\rm \odot}$
(Table\,3), were adopted from
Grevesse \& Sauval (1998) with the exception of oxygen for which 
the abundance of Asplund et al. (2004) was adopted. 
A MARCS model with the parameters $T_{\rm eff}=5780$\,K, $\log g=4.44$\,dex,
[Fe/H]\,$=0.00$\,dex and $\xi_{\rm t}=1.15$\,km/s was used.
For the oxygen triplet lines, corrections for NLTE ($-0.22$\,dex) were
applied when determining the $gf$ values.

The choice of a differential analysis relative to the Sun may not be optimal in 
our case. The stars depart systematically from the Sun in one important 
respect -- they are more metal-poor and therefore their lines from heavy 
elements are generally weaker. Also, the choice of a MARCS model is not 
obvious. True enough, errors may cancel if one chooses the Solar model from 
the same grid as for the programme stars. However, this cancellation may be 
only partial since the programme stars are systematically different. Therefore, 
arguments for a semi-empirical solar model instead can be raised, c.f.
Sect.\,\ref{SHMmodel}.
\begin{figure*}
\centering
\resizebox{16cm}{!}{\includegraphics*{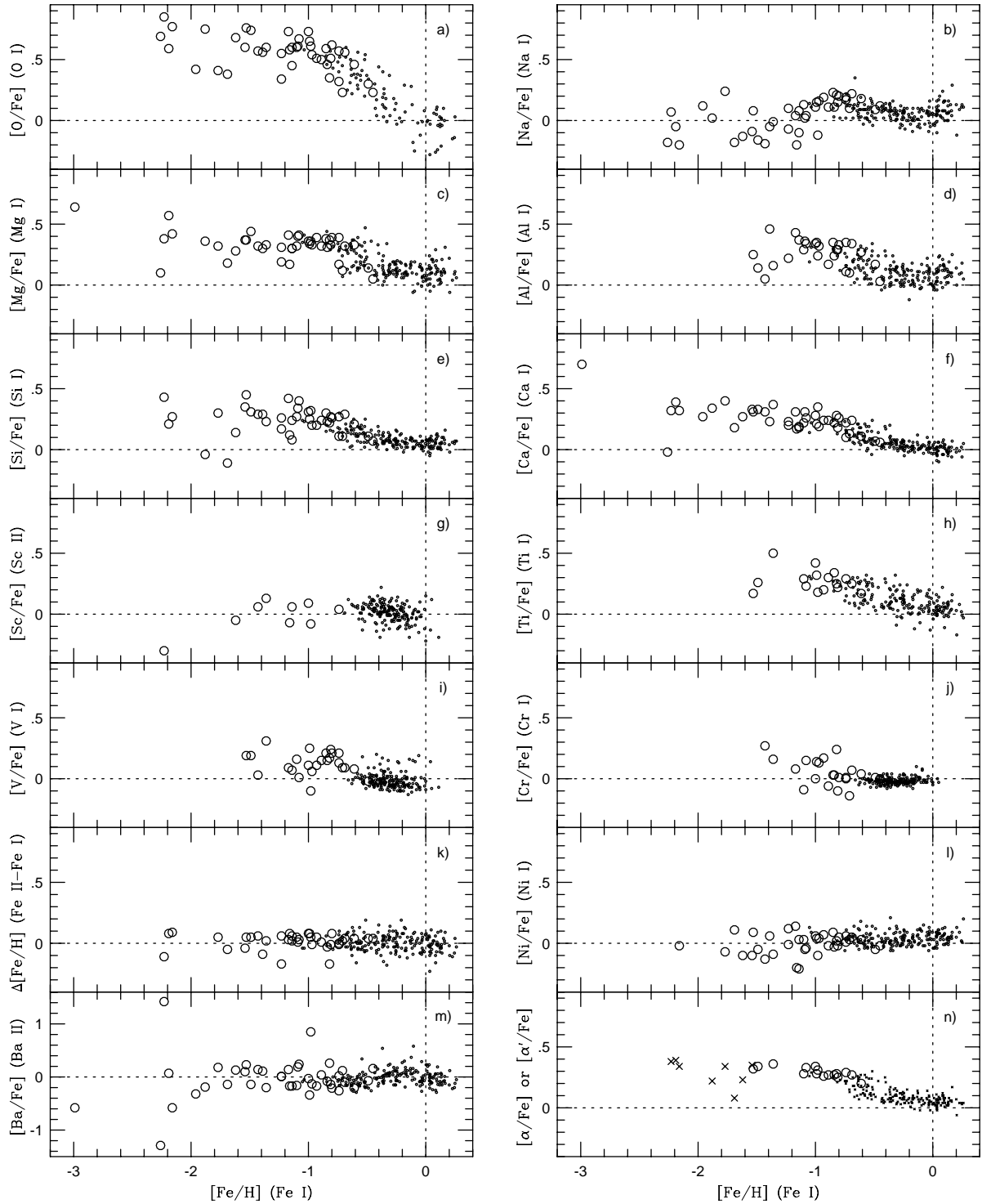}}
\caption[]{\label{Fresults}
The abundances relative to iron, [X/Fe]$_{\rm Fe I}$ for all species in the
programme as a function of the iron abundance with addition of the data of
Edvardsson et al. (1993) as small dots.
For species where Edvardsson et al. have no data (Sc\,{\sc i}, V\,{\sc i} and
Cr\,{\sc i}) the data of Reddy et al. (2003) is shown instead.
Note the different ordinate scale in panel m).
In panel n) the circles denote
[$\alpha$/Fe]\,$=\frac{1}{4}({\rm [Mg/Fe]}+{\rm [Si/Fe]}+
{\rm [Ca/Fe]}+{\rm [Ti/Fe]})$ and the crosses
[$\alpha$'/Fe]\,$=\frac{1}{3}({\rm [Mg/Fe]}+{\rm [Si/Fe]}+{\rm [Ca/Fe]})$ for stars
with no Ti abundance determinations.  }
\end{figure*}

The solar equivalent widths were measured from a computer readable version of
the atlas of Kurucz et al. (1984).
The lines were fitted twice with the programme package IRAF:
with a Gaussian fit and alternately by a numerical integration.
The final width was set as a weighted mean of these values, with the weighting
determined by visual inspection.

When determining the $gf$ value of the Ba\,{\sc ii} 6141.713\,\AA\ line, which
has an Fe\,{\sc i} blend, we first scaled the iron line $gf$ value to two other
lines of the same multiplet, 6232.5 and 6302.5\,\AA\, which were fitted to the
solar flux atlas of Kurucz et al. (1984).
The relative scaling was based on the $gf$ values listed by Bard et al. (1991),
and the fitting of the two lines resulted in increased $\log gf$ values by
$+0.02$ and $+0.10$\,dex, respectively.
The adopted $\log gf$ value of the Fe\,{\sc i} is $-1.40$; 0.06\,dex higher than
given by Bard et al.
Synthesis of the Ba\,{\sc ii} line including the (negligible) hfs and IS
splitting and the Fe\,{\sc i} line to the solar atlas, indicated that the Ba
and Fe lines do not differ in central wavelengths by more than 0.005\,\AA\. 
Therefore these wavelengths were set to be identical.
Finally, requiring the total equivalent width of the blend to be equal to our
measured solar equivalent width resulted in a $\log gf$ for the Ba\,{\sc ii}
line of $-0.13$, which is 1.1\,$\sigma$ lower than the NBS value of $-0.077$
(Wiese \& Martin 1980).

\section{Results}
\label{Sresults}
The results of the abundance analysis are given in Table\,\ref{Tresults} and 
displayed in Fig.\,\ref{Fresults}.
Here we have added the abundances of disk stars, obtained with very similar data
and methods.
A number of different morphological patterns are shown 
in the figure, and many of these have been found and discussed by others.  
We recognise the ``$\alpha$-element behaviour'' of Mg, Si, Ca, and possibly of 
Ti, with a rise of [X/Fe] as one proceeds towards the more metal-poor
stars, and a levelling off to a plateau for [Fe/H]\,$<-$1.
It is not clear, however, that this plateau level is strictly horizontal, a
question which will be discussed in Sect.\,\ref{Sdiscussion}.
Also O shows a similar behaviour.
We see an ``iron-like'' behaviour of Ni and possibly Cr, with [X/Fe] staying
close to solar also for metal-poor stars.
We find the ``odd-Z element'' Al to resemble the 
even-Z alpha elements (cf Si!), while the odd Na behaves differently.
We find that Sc, V and Ba do not show any clear rise relative to Fe with
decreasing [Fe/H] and that the [Ba/Fe] ratio (based on only one blended feature)
shows a substantial scatter, seemingly caused by some pronounced outliers.

In all panels of Fig.\,\ref{Fresults} where our data overlap with the disk star
data there is a smooth transition between the data sets.

One important question is if the scatter around the mean trends in
Fig.\,\ref{Fresults} are real or caused by errors in the analysis.  E.g., is 
the increased scatter when we proceed to stars with [Fe/H]\,$<-1.0$, visible for 
most elements, due to a real spread in abundances or caused by weaker and fewer
measurable lines, lower S/N or more uncertain fundamental parameters?
Do the differences in scatter, ranging from 0.03\,dex (in [Ca/Fe]) to above
0.1\,dex (e.g. in [O/Fe], [Al/Fe], [Ba/Fe]) around the mean relations tell us
anything about nucleosynthesis?
Is any of the scatter in the alpha elements [Si/Fe] and [Mg/Fe] real, in view of
the smaller scatter in [Ca/Fe], at a given [Fe/H]? 
The discussion of such questions will be postponed until after the error
discussion in Sect.\,\ref{Sconschecks} and a detailed
comparison with results from other studies in Sect.\,\ref{Sdiscussion}. 

One additional feature of Fig.\,\ref{Fresults} should be pointed out -- the 
nice agreement in iron abundances as derived from the Fe\,{\sc i} lines and the
(single!) Fe\,{\sc ii} line.
This agreement will also be further discussed in Sect.\,\ref{Sdiscussion}.
\setcounter{table}{3}
\begin{table*}
\begin{scriptsize}
\caption[]{
\label{Tresults}
Spectroscopic metallicities, [Fe/H]$_{\rm Fe\,I}$, and abundances relative to
iron, [X/Fe]$_{\rm Fe\,I}$, derived for 13 neutral and ionized species X.
For oxygen, NLTE corrections have been applied.
[$\alpha$/Fe]=$\frac{1}{4}({\rm [Mg/Fe]}+{\rm [Si/Fe]}+{\rm [Ca/Fe]}+{\rm [Ti/Fe]})$
and [$\alpha$'/Fe]=$\frac{1}{3}({\rm [Mg/Fe]}+{\rm [Si/Fe]}+{\rm [Ca/Fe]})$.}
\end{scriptsize}
\begin{flushleft} 
\begin{tiny}
\begin{tabular*}{18.0cm}{lrrrrrrrrrrrrrrr}
HD & [Fe/H] & O\,{\sc i} & Na\,{\sc i} & Mg\,{\sc i} &
Al\,{\sc i} & Si\,{\sc i} & Ca\,{\sc i} & Sc\,{\sc ii} & Ti\,{\sc i} &
V\,{\sc i} & Cr\,{\sc i} & Fe\,{\sc ii} & Ni\,{\sc i} & Ba\,{\sc ii} &
$\alpha$/$\alpha'$ \\
\hline
17072  & $-$0.98 &    0.61 & $-$0.12 &    0.36 &    0.24 &    0.32 &    0.35 & $-$0.08 &    0.18 & $-$0.10 &         &    0.05 & $-$0.10 &    0.59 &    0.31/0.34 \\
25704  & $-$0.97 &    0.54 &    0.16 &    0.33 &    0.32 &    0.20 &    0.19 &         &         &    0.06 &    0.13 & $-$0.01 &    0.04 & $-$0.24 &        /0.24 \\
49301  & $-$0.82 &    0.35 &    0.21 &    0.37 &    0.29 &    0.22 &    0.28 &         &    0.25 &    0.17 &    0.24 & $-$0.17 &    0.02 &    0.04 &    0.28/0.29 \\
59374  & $-$0.93 &    0.51 &    0.19 &    0.39 &         &    0.20 &    0.24 &         &    0.20 &    0.11 &    0.17 &    0.05 &    0.07 & $-$0.26 &    0.26/0.28 \\
61902  & $-$0.71 &    0.23 &    0.10 &    0.12 &    0.10 &    0.11 &    0.13 &         &         &    0.09 & $-$0.14 &    0.02 &    0.03 & $-$0.03 &        /0.12 \\
63598  & $-$0.84 &    0.46 &    0.11 &    0.31 &    0.24 &    0.23 &    0.22 &         &    0.34 &    0.15 &    0.03 & $-$0.03 & $-$0.03 & $-$0.19 &    0.28/0.25 \\
76932  & $-$0.89 &    0.50 &    0.11 &    0.32 &    0.17 &    0.24 &    0.24 &         &    0.30 &    0.15 & $-$0.06 &    0.01 & $-$0.02 & $-$0.11 &    0.27/0.27 \\
78747  & $-$0.80 &    0.62 &    0.20 &    0.39 &    0.33 &    0.26 &    0.26 &         &         &    0.21 &    0.01 &    0.08 &    0.05 & $-$0.28 &        /0.30 \\
79601  & $-$0.74 &    0.57 &    0.19 &    0.39 &    0.35 &    0.27 &    0.22 &         &    0.29 &    0.21 &    0.01 &    0.00 &    0.06 & $-$0.34 &    0.29/0.29 \\
80863  & $-$0.61 &    0.46 &    0.18 &    0.33 &    0.27 &    0.21 &    0.11 &         &    0.17 &    0.08 &    0.04 &    0.04 &    0.03 & $-$0.27 &    0.20/0.22 \\
84937  & $-$1.96 &    0.42 &    0.12 &         &         &         &    0.27 &         &         &         &         &         &         & $-$0.44 &              \\
91121  & $-$1.08 &    0.67 &    0.04 &    0.41 &    0.34 &    0.40 &    0.26 &         &    0.23 &    0.01 &    0.15 &    0.03 & $-$0.04 &    0.07 &    0.33/0.36 \\
94028  & $-$1.39 &    0.56 & $-$0.05 &    0.30 &    0.46 &    0.29 &    0.23 &         &         &         &         & $-$0.09 &    0.06 & $-$0.04 &        /0.27 \\
97320  & $-$1.23 &    0.55 &    0.10 &    0.31 &    0.22 &    0.26 &    0.23 &         &         &         &         &    0.06 &    0.12 & $-$0.14 &        /0.27 \\
99682  & $-$0.45 &    0.23 &    0.12 &    0.05 &    0.03 &    0.07 &    0.06 &         &         &    0.00 &    0.00 &    0.04 & $-$0.02 &    0.02 &        /0.06 \\
101976 & $-$0.49 &    0.30 &    0.09 &    0.14 &    0.17 &    0.11 &    0.07 &         &         & $-$0.02 &    0.01 &    0.04 & $-$0.05 & $-$0.17 &        /0.11 \\
102200 & $-$1.23 &    0.44 & $-$0.07 &    0.19 &         &    0.17 &    0.20 &         &         &         &         & $-$0.17 & $-$0.01 & $-$0.14 &        /0.19 \\
108317 & $-$2.19 &    0.59 & $-$0.05 &    0.57 &         &    0.21 &    0.39 &         &         &         &         &    0.08 &         & $-$0.17 &        /0.39 \\
111777 & $-$0.85 &    0.59 &    0.23 &    0.38 &    0.35 &    0.30 &         &         &         &    0.21 &    0.03 &         &    0.09 &         &              \\
116064 & $-$1.88 &    0.75 &    0.02 &    0.36 &         & $-$0.04 &    0.34 &         &         &         &         &         &         & $-$0.32 &        /0.22 \\
120559 & $-$0.99 &    0.65 &    0.15 &    0.34 &    0.35 &    0.25 &    0.21 &         &    0.32 &    0.25 &    0.14 &    0.08 &    0.04 & $-$0.32 &    0.28/0.27 \\
122196 & $-$1.69 &    0.38 & $-$0.18 &    0.18 &         & $-$0.11 &    0.18 &         &         &         &         & $-$0.05 &    0.11 & $-$0.29 &        /0.08 \\
124785 & $-$0.69 &    0.56 &    0.22 &    0.32 &    0.34 &    0.29 &    0.24 &         &    0.25 &    0.09 &    0.07 &    0.04 &    0.05 & $-$0.21 &    0.27/0.28 \\
126793 & $-$0.81 &    0.51 &    0.15 &    0.33 &    0.29 &    0.27 &    0.18 &         &    0.22 &    0.24 & $-$0.10 & $-$0.01 & $-$0.02 & $-$0.24 &    0.25/0.26 \\
128279 & $-$2.16 &    0.77 & $-$0.20 &    0.42 &         &    0.27 &    0.32 &         &         &         &         &    0.09 & $-$0.02 & $-$0.74 &        /0.34 \\
132475 & $-$1.53 &    0.76 &    0.08 &    0.37 &    0.25 &    0.45 &    0.31 &         &    0.17 &    0.19 &         &    0.05 &    0.09 &    0.06 &    0.32/0.38 \\
140283 & $-$2.26 &    0.69 & $-$0.18 &    0.10 &         &         & $-$0.02 &         &         &         &         &         &         & $-$1.11 &              \\
142945 & $-$1.00 &    0.73 &    0.11 &    0.36 &    0.34 &    0.31 &    0.28 &    0.09 &    0.42 &    0.11 &    0.00 &    0.08 &    0.06 & $-$0.14 &    0.34/0.32 \\
144450 & $-$1.17 &    0.73 &    0.04 &    0.41 &    0.43 &    0.42 &    0.31 &         &         &    0.09 &    0.08 &    0.03 &    0.14 &    0.00 &        /0.38 \\
145417 & $-$1.36 &    0.60 & $-$0.01 &    0.33 &    0.16 &    0.23 &    0.37 &    0.13 &    0.50 &    0.31 &    0.16 &    0.02 & $-$0.09 & $-$0.16 &    0.36/0.31 \\
160617 & $-$1.77 &    0.41 &    0.24 &    0.32 &         &    0.30 &    0.40 &         &         &         &         &    0.05 & $-$0.07 &    0.01 &        /0.34 \\
166913 & $-$1.54 &    0.60 & $-$0.09 &    0.37 &         &    0.35 &    0.33 &         &         &         &         & $-$0.04 & $-$0.10 & $-$0.07 &        /0.35 \\
188510 & $-$1.62 &    0.68 & $-$0.13 &    0.28 &         &    0.14 &    0.27 & $-$0.05 &         &         &         &         & $-$0.10 & $-$0.25 &        /0.23 \\
193901 & $-$1.16 &    0.58 & $-$0.20 &    0.17 &         &    0.12 &    0.17 & $-$0.07 &         &         &         &    0.08 & $-$0.20 & $-$0.25 &        /0.15 \\
194598 & $-$1.14 &    0.45 & $-$0.10 &    0.30 &         &    0.08 &    0.18 &    0.06 &         &         &         &    0.06 & $-$0.21 &         &        /0.19 \\
196892 & $-$1.09 &    0.61 &    0.02 &    0.40 &    0.36 &    0.34 &    0.31 &         &         &         &         &    0.01 & $-$0.05 &    0.04 &        /0.35 \\
196944 & $-$2.23 &    0.85 &    0.07 &    0.38 &         &    0.43 &    0.32 & $-$0.30 &         &         &         & $-$0.11 &         &    1.14 &        /0.38 \\
199289 & $-$1.10 &    0.60 &    0.13 &    0.32 &    0.29 &    0.27 &    0.22 &         &    0.29 &    0.16 & $-$0.09 &    0.05 &    0.03 & $-$0.28 &    0.28/0.27 \\
200654 & $-$2.99 &         &         &    0.64 &         &         &    0.70 &         &         &         &         &         &         & $-$0.73 &              \\
201891 & $-$1.14 &    0.60 &    0.08 &    0.30 &    0.37 &    0.24 &    0.19 &         &         &    0.07 &         &    0.02 &    0.03 & $-$0.30 &        /0.24 \\
203608 & $-$0.74 &    0.32 &    0.17 &    0.17 &    0.11 &    0.11 &    0.10 &    0.04 &         &    0.13 &    0.00 & $-$0.01 &    0.01 & $-$0.11 &        /0.13 \\
211998 & $-$1.49 &    0.74 & $-$0.16 &    0.44 &    0.14 &    0.31 &    0.33 &         &    0.26 &    0.19 &         &    0.05 & $-$0.05 & $-$0.29 &    0.34/0.36 \\
213467 & $-$1.43 &    0.57 & $-$0.19 &    0.32 &    0.05 &    0.29 &    0.31 &    0.06 &         &    0.03 &    0.27 &    0.06 & $-$0.13 & $-$0.03 &        /0.30 \\
\hline
\end{tabular*}
\end{tiny}
\end{flushleft}
\end{table*}

\section{Consistency checks and error estimates}
\label{Sconschecks}
\subsection{Abundance dependence on individual lines}
\label{Scheckslines}
\begin{figure}
\centering
\resizebox{\hsize}{!}{\includegraphics*{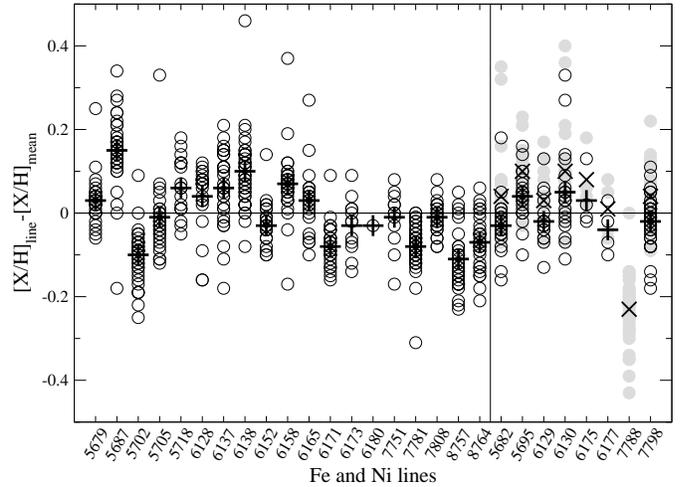}}
\caption[]{\label{Fcheckslines}
A consistency check of the abundances of iron (left) and nickel (right).
The open black circles mark the measurements for each line for each
star and the thick ($+$):s mark the mean for individual lines. For nickel the
filled grey circles and the thick ({\bf x}):s mark the measurements with the
line 7788\,{\AA} included. This line was later removed from the analysis
due to suspected blending.}
\end{figure}

A check was performed to see if any individual lines of the elements 
represented by numerous lines in our spectra tend to give systematically 
different abundances from the rest of the lines. The relative abundances 
[X/H]$_{\rm line}-$[X/H]$_{\rm mean}$ were thus calculated for all iron and 
nickel lines included in the programme, see Fig.\,\ref{Fcheckslines}. Some 
lines did give systematically different abundances from the mean abundance of
the species, and the Ni line 7788\,{\AA} was removed from the abundance 
determinations altogether.
The reason for the deviating abundances from individual lines 
could be due to hidden blends in the Sun, and therefore erroneous $\log gf$ 
values, or hidden blends or departures from LTE in the stellar spectra depending 
on the parameters. 

Some silicon lines (6125\,{\AA}, 6142\,{\AA}, 6145\,{\AA} and 6155\,{\AA}) come
from double, not single, electron transitions in the atoms.
We have checked whether these lines behave differently compared to the rest, but
they do not differ more than expected statistically.
\begin{table}
\begin{scriptsize}
\caption[]{\label{Tabundchange}
The table show the effects on the derived logarithmic abundances relative to
hydrogen, $\Delta$[X/H]=[X/H]$_{\rm changed}$-[X/H]$_{\rm normal}$, of changes
in effective temperature (K), surface gravity (dex), metallicity (dex) and
microturbulence (km/s) on the mean abundance for all 43 stars.
The last two columns show the abundance change if the $\log gf$-values were
calculated with only the Uns\"old approximation, see Sect.\,\ref{Slinebroadening}
or based on the Holweger \& M\"uller model, see Sect.\,\ref{SHMmodel}. }
\begin{flushleft} 
\begin{tabular*}{8.8cm}{@{\extracolsep\fill}l|rrrrrr}
Ion & \multicolumn{6}{c}{Mean abundance change for all 43 stars} \\
& $T_{\rm eff}$ & $\log g$   & [Me/H]     & $\xi_{\rm t}$ & $\log gf$ & $\log gf$ \\
& $+$100\,K     & $+$0.2     & $+$0.5     & $+$0.5        & Uns\"old  & HM \\
& $-$100\,K     & $-$0.2     & $-$0.5     & $-$0.5        & only      & model \\
\hline
O\,{\sc I}   &  $-$0.08 &     0.06 &     0.00 &  $-$0.03 &  $-$0.01 & -0.03\\ 
             &     0.08 &  $-$0.06 &     0.01 &     0.02 & & \\ 
Na\,{\sc I}  &     0.05 &     0.00 &     0.01 &  $-$0.01 &  $-$0.02 & -0.12\\ 
             &  $-$0.05 &     0.01 &  $-$0.01 &     0.01 & & \\ 
Mg\,{\sc I}  &     0.04 &  $-$0.01 &     0.02 &  $-$0.01 &    $^*$ & -0.11\\ 
             &  $-$0.04 &     0.02 &  $-$0.02 &     0.01 & & \\ 
Al\,{\sc I}  &     0.03 &  $-$0.01 &     0.01 &  $-$0.01 &     0.00 & -0.09\\ 
             &  $-$0.03 &     0.01 &  $-$0.01 &     0.01 & & \\ 
Si\,{\sc I}  &     0.03 &     0.01 &     0.02 &  $-$0.01 &  $-$0.02 & -0.09\\ 
             &  $-$0.02 &  $-$0.01 &  $-$0.01 &     0.01 & & \\ 
Ca\,{\sc I}  &     0.06 &  $-$0.02 &     0.02 &  $-$0.04 &  $-$0.03 & -0.15\\ 
             &  $-$0.07 &     0.02 &  $-$0.03 &     0.05 & & \\ 
Sc\,{\sc II} &     0.02 &     0.08 &     0.08 &  $-$0.01 &    $^*$ & -0.06\\ 
             &  $-$0.02 &  $-$0.08 &  $-$0.05 &     0.02 & & \\ 
Ti\,{\sc I}  &     0.09 &     0.00 &     0.00 &     0.00 &     0.00 & -0.10\\ 
             &  $-$0.10 &     0.00 &  $-$0.01 &     0.00 & & \\ 
V\,{\sc I}   &     0.10 &     0.00 &     0.01 &     0.00 &     0.00 & -0.11\\ 
             &  $-$0.10 &     0.00 &  $-$0.01 &     0.01 & & \\ 
Cr\,{\sc I}  &     0.05 &     0.00 &     0.00 &     0.00 &     0.04 & -0.10\\ 
             &  $-$0.06 &     0.00 &     0.00 &     0.01 & & \\ 
Fe\,{\sc I}  &     0.07 &     0.00 &     0.02 &  $-$0.03 &  $-$0.04 & -0.15\\ 
             &  $-$0.07 &     0.00 &  $-$0.02 &     0.04 & & \\ 
Fe\,{\sc II} &  $-$0.01 &     0.08 &     0.05 &  $-$0.01 &    $^*$ & -0.06\\ 
             &     0.02 &  $-$0.07 &  $-$0.03 &     0.02 & & \\ 
Ni\,{\sc I}  &     0.05 &     0.01 &     0.02 &  $-$0.01 &     0.02 & -0.12\\ 
             &  $-$0.05 &  $-$0.01 &  $-$0.02 &     0.01 & & \\ 
Ba\,{\sc II} &     0.07 &     0.03 &     0.09 &  $-$0.20 &     0.08 & -0.15\\ 
             &  $-$0.06 &  $-$0.04 &  $-$0.06 &     0.22 & & \\ 
\hline
\multicolumn{7}{l}{$^*$ These species have only lines calculated with the Uns\"old} \\
\multicolumn{7}{l}{approximation from the start.}
\end{tabular*}
\end{flushleft}
\end{scriptsize}
\end{table}

\subsection{Errors in the stellar effective temperatures}
\label{Stemperror}
The uncorrected ($b-y$) colour, as well as other uncorrected indices were used
for estimating stellar effective temperatures. This 
should not affect the result much because the interstellar reddening, derived 
from the observed ($b-y$) minus the reddening free ($b-y$)$_0$ according to 
Olsen (1988), were found to be negligible for all stars except \object{HD\,132475} 
for which the colour excess was $E(b-y)=0.034$.
This corresponds to an effective temperature error of 180\,K.
Hakkila et al. (1997) have constructed a model for interstellar reddening in the 
solar neighbourhood based on published results from 
large-scale surveys of interstellar extinction.
Using that model we find reddenings of typically $E(b-y)=$\,0.00 to 0.015\,mag,
the median value being 0.010.
For a few stars we find values extending beyond 0.03\,mag, namely
HD\,126793, 132475, 144450, 160617, 193901 and 196944.
The largest values appear for \object{HD\,126793} ($E(b-y)=0.046$),
\object{HD\,144450} (0.052) and \object{HD\,196944} (0.065).
For \object{HD\,132475} the Hakkila et al. model gives $E(b-y) = 0.025$, i.e. somewhat
smaller than the value from the H$_\beta$ photometry.
However, the mean error in the estimates from the model is typically 0.05\,mag.
We conclude that reddenings of typically 0.01\,mag
may affect many of our programme stars, leading to underestimates
of effective temperature by about 50\,K and of metal abundances 
of about 0.03\,dex. For a few stars, like \object{HD\,132475} there 
may be more serious underestimates, by even somewhat more than 0.1\,dex.   

A test was made to investigate the impact of the temperature errors.
Alonso et al. (1996) estimate a standard deviation of 110\,K in the effective
temperatures.
Here the parameters were changed $\pm$100\,K for all stars.
The changes of the chemical abundances were at a mean
$\pm0.04$\,dex, but for O\,{\sc i}, Ca\,{\sc i}, Ti\,{\sc i}, V\,{\sc i},
Fe\,{\sc i} and Ba\,{\sc ii} the mean abundances changed by more than
$0.07$\,dex, see Table\,\ref{Tabundchange}.
So, errors in the effective temperatures have considerable impact on the result. 

A frequently used consistency test on the effective temperatures is to check 
if the derived abundances are dependent on the excitation energies for the 
lower states of the lines.
The level populations depend on the temperatures 
through the Boltzmann distribution in the LTE approximation.
The numerous lines of Fe\,{\sc i} and Ni\,{\sc i} with wide ranges both in
excitation energy and equivalent widths were selected for the consistency
checks. 
The use of both species together is motivated by the 
similarities between the elements and the gain in statistics with as many lines 
as possible. 
\begin{figure}
\centering
\resizebox{\hsize}{!}{\includegraphics*{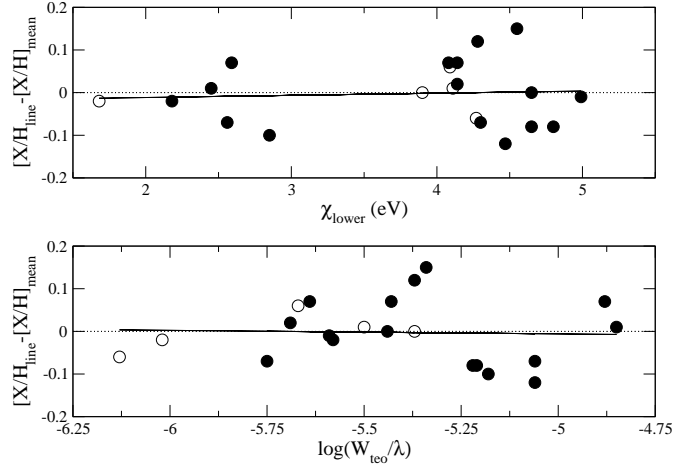}}
\caption[]{\label{Fconsexample}
An example of consistency diagrams for an individual star, in this case
\object{HD\,126793}. Filled circles represent iron lines and open circles nickel
lines.
Linear fits were made to the data for individual stars, and the derived slopes
were studied as functions of stellar parameters, e.g., in Fig.\,\ref{Fconstemp}.
}
\end{figure}
\begin{figure}
\centering
\resizebox{\hsize}{!}{\includegraphics*{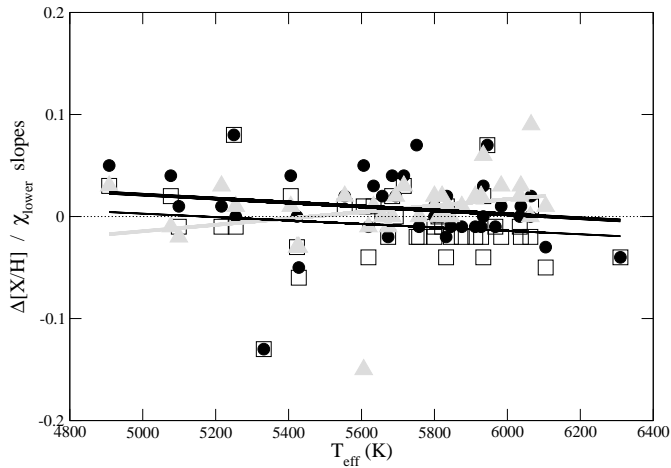}}
\caption[]{\label{Fconstemp}
A consistency check for the effective temperatures.
Filled circles and thick black line: all lines included.
Squares and black line: the lines 5687 and 8757\,{\AA} were removed.
Grey triangles and grey line: the lines at
5687, 5701, 5717, 6136, 6137, 6157, 6170, 7780, 8757, and 8763\,{\AA}
were removed.}
\end{figure}

For each star included in the tests a diagram was plotted which displayed the 
abundance from each line minus the derived mean abundance for the star, 
[X/H]$_{\rm line}-$[X/H]$_{\rm mean}$, versus the excitation energy of the 
lower level of the line, $\chi_{\rm lower}$, see Fig.\,\ref{Fconsexample} top 
panel. To all diagrams a linear fit was sought and these slopes were then 
plotted in a new diagram with the effective temperatures of the stars on the 
abscissa, see Fig.\,\ref{Fconstemp}. 

The check shows that the test is very dependent on the lines employed in the 
test.
If all lines except Ni\,{\sc i} 7788\,\AA\ are included,
the test suggests that the adopted effective temperatures are about 100\,K too
low for the stars in the lower temperature region and about 25\,K too
high at high temperatures (thick black line).
However, if lines deviating by more than 0.1\,dex from the mean in
Fig.\,\ref{Fcheckslines} are removed (removal of the iron lines at
5687 and 8757\,{\AA}), the temperatures of the low-temperature
stars seem underestimated by some 20\,K, while the warmer stars
have their temperatures overestimated by about 120\,K
(thin black line).
If spectral lines deviating by more than 0.05\,dex are removed (removal of the
iron lines at 5687, 5701, 5717, 6136, 6137, 6157, 6170, 7780, 8757 and
8763\,{\AA}), then adopted effective temperatures
of the low-temperature stars are about 70\,K too high and high
temperature stars have temperatures about 130\,K too low (thick grey line).
The conclusion is that this test, assuming that the LTE approximation is valid,
suggests that the errors in temperature are not larger than 150\,K.
For a few stars the temperatures may, however, be underestimated by even more,
due to reddening.
More precise conclusions cannot be drawn.

\subsection{Errors in microturbulence parameters}
\label{microerror}
A test was made to investigate the impact of the 
microturbulence errors. The parameters were changed $\pm0.5$\,km/s for all stars. 
The average change of the mean abundances were only $\mp0.03$\,dex. An 
exception was barium, for which the change was $\mp0.21$\,dex.

We have also checked if the adopted value of 1.5\,km/s for all stars (except for 
\object{HD\,196944}) give consistent results by investigating whether the line
strengths, $\log\left(W_{\rm theo}/\lambda\right)$, and the relative abundances
of the lines, [X/H]$_{\rm line}-$[X/H]$_{\rm mean}$, are correlated.
$W_{\rm theo}$ represents the width the line would have if the mean abundance
from all lines is adopted, following Magain (1984).
If the assumed microturbulence parameter, and the underlying line-formation
theory are correct, the relative abundances derived from the individual lines
would be independent of line strength. 

The test was performed in a way similar to that for the effective temperature,
see previous section. A diagram was plotted for each star with the relative 
abundance, [X/H]$_{\rm line}-$[X/H]$_{\rm mean}$, on the ordinate and the 
theoretical line strength, $\log\left(W_{\rm theo}/\lambda\right)$, on the 
abscissa, see the lower panel of Fig.\,\ref{Fconsexample}. To all figures a 
linear fit was sought.

The results show again that this kind of test is heavily dependent on which
lines are included, although one may make the conclusion that the 
microturbulences are slightly overestimated in general and not optimal for some 
stars. This has however, a very little impact on the derived mean elemental 
abundances, with the exception of barium, as discussed in Sect.\,\ref{Sbarium}.

\subsection{Errors in metallicities}
\label{Smetalerror}
In the present study, the iron abundance, [Fe/H], as derived spectroscopically
from lines of Fe\,{\sc i}, has been used to represent the over-all metallicity
in the model atmosphere used for the star.
The effects of errors in the metallicities in the analysis 
have been tested in the following way: 

The metallicities of the model atmospheres were changed $\pm0.5$\,dex for all
stars.
The change of the mean abundances were on average $\pm0.02$\,dex, but
higher for Sc\,{\sc ii}, Fe\,{\sc ii} and Ba\,{\sc ii}, see
Table\,\ref{Tabundchange}.
The conclusion is that errors in the metallicities do not have a large impact.

\begin{figure}
\centering
\resizebox{\hsize}{!}{\includegraphics*{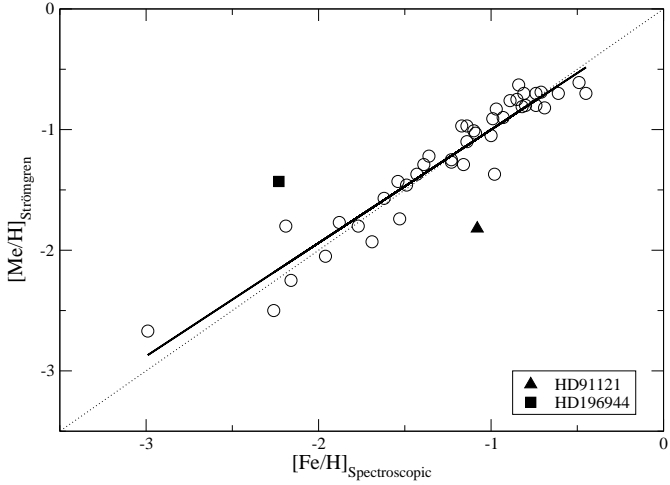}}
\caption[]{\label{FMeFe}
Comparison of derived iron abundances, [Fe/H], with metallicities from
Str\"{o}mgren photometry obtained from the calibration of
Schuster \& Nissen (1989).
The solid line shows equation\,\ref{EMeFe} and the dotted line indicates
the 1-to-1 relation.
}
\end{figure}

The derived iron abundances were compared with metallicities given by the
Schuster \& Nissen (1989) calibration of Str\"{o}mgren photometry
see Fig.\,\ref{FMeFe}.
(Note that the correction to Magain's (1987, 1989) [Fe/H] scale has not been
applied -- this scale was found by Gosset \& Magain (1993) to be affected by
a non-linear detector.)
The derived abundances differed from the photometric ones by
$-0.01$\,dex on average with a standard deviation of 0.25\,dex, a maximum
difference of 1.01\,dex (\object{HD\,91121}, with photometry from a separate
source) and a minimum difference of $-0.58$\,dex (\object{HD\,196944},
supergiant).
If we remove the 2 outliers,
the mean difference does not change and the equation of the black line is
\begin{eqnarray}
[{\rm Me/H}]_{\rm Stromgren}=-0.06+0.94\,[{\rm Fe/H}]_{\rm Spectroscopic}
\label{EMeFe}
\end{eqnarray}
with a scatter of $0.15$\,dex.

\subsection{Errors in surface gravities}
\label{loggerror}
The gravities derived from Hipparcos parallaxes were estimated to have an
average standard deviation error of $0.09$\,dex.
These errors were derived assuming an error of 70\,K for the effective
temperatures of the stars.
For the gravity parameter of the star \object{HD\,196944}
Za\u{c}s et al. (1998), claim an error interval of $\pm0.3$\,dex.

\begin{figure}
\centering
\resizebox{\hsize}{!}{\includegraphics*{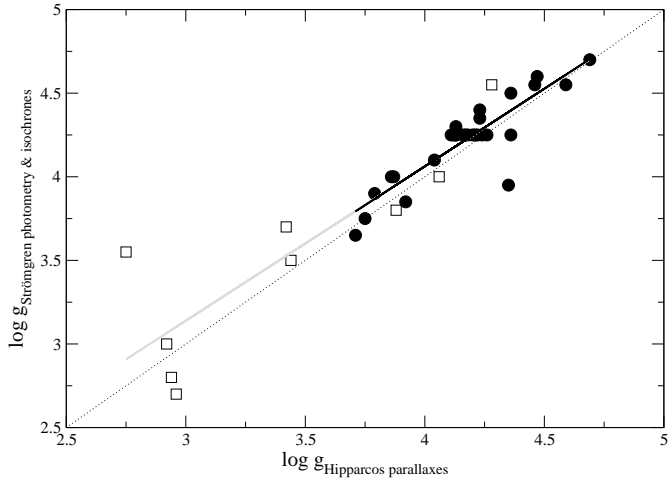}}
\caption[]{
Comparison between $\log g$ from Hipparcos and
Str\"{o}mgren photometry respectively, see text.
The filled circles and black line represent stars and fits with reliable data
from Str\"{o}mgren photometry, and squares more uncertain measurements.
The grey line shows a fit to all data. The dotted line indicates a 1-to-1
relation.
\label{Floggcomp} }
\end{figure}

Our neglect of interstellar absorption may make our assumed reddening-free
$V$ magnitudes too large.
Using the result from Sect.\,\ref{Stemperror} above, we estimate that typical
$A_V$ values hardly exceed 0.04\,mag for most stars, and not 0.4\,mag for any
of them.
This corresponds to overestimates in $\log g$ by 0.02 to 0.16\,dex,
respectively.
We have compared our gravity estimates with those obtained from Str\"{o}mgren
photometry and the isochrones of VandenBerg \& Bell (1985).
The results agree very well with those obtained here 
from the Hipparcos parallaxes. We thus find a relation
\begin{eqnarray}
\log g {\rm (isochrones)}= -0.37 + 0.92 \log g{\rm (Hipparcos)}
\label{EIsoHip}
\end{eqnarray}
with a standard deviation of 0.18\,dex.
In fact, if the coolest stars with $\log g < 3.5$\,dex are excluded,
the standard deviation diminishes to 0.11\,dex
(cf. Fig.\,\ref{Floggcomp}).
From these considerations, it seems reasonable to adopt an uncertainty on
the order of 0.2\,dex for our estimates of surface gravity parameters. 
The effects of these errors on the final element abundances
have been explored.
The change of the elemental abundances were only $\pm0.02$\,dex,
but for the elements O\,{\sc i}, Sc\,{\sc ii}, Fe\,{\sc ii} and
Ba\,{\sc ii}, the changes were larger, from $\pm0.04$ to $\pm0.08$\,dex,
respectively, see Table\,\ref{Tabundchange}.

\subsection{Errors in the oscillator strengths}
\subsubsection{Pressure broadening}
\label{Slinebroadening}
If the Uns\"old approximation was used for all lines when fitting the solar
equivalent widths, the $\log gf$ values became on the mean
$0.03$\,dex higher with a maximum of $+0.26$\,dex.
The mean abundances of the stellar sample changed very little, see
Table\,\ref{Tabundchange}, but for stars with few measured lines of some
element the effect may be important.

\subsubsection{Solar equivalent widths and $\log gf$ values}
The stellar and solar equivalent widths
were estimated through a weighted mean of a Gaussian fit and pure integration.
With this method the solar 
equivalent widths became larger than the ones used in the investigation of disk 
dwarf stars by Edvardsson et al. (1993); They only used 
Gaussian fits to their lines, and therefore parts of the line wings were not 
included. The mean difference was 2.2\,m{\AA} (3.3\%), with a maximum of 
8.6\,m{\AA} (9.9\%). 
However, Edvardsson et al. systematically used the Uns\"old approximation while
our application of the O'Mara, Anstee \& Barklem data lower the
resulting $\log gf$ values, counteracting the effects of our higher solar
equivalent widths.
Except for the O\,{\sc i} triplet lines, for which Edvardsson et al. did not
make any NLTE corrections, the mean difference in $\log gf$
(us$-$Edvardsson et al.) is 0.02\,dex with a standard deviation of 0.05\,dex
for the 28 lines in common.

The error in the $gf$ values deduced for the blended Ba\,{\sc ii} line at
6141\,\AA\ depends on the strength adopted for the blending Fe\,{\sc i}
line.
If the latter line is totally neglected, we find the $gf$ value for the
Ba\,{\sc ii} line to increase to $\log gf = 0.09$, and the typical Ba
abundances for the programme stars to decrease by 0.15\,dex.
An increase of the $gf$ value of the Fe\,{\sc i} line by 0.2\,dex cannot be
excluded, which leads to $\log gf$(Ba\,{\sc ii})\,$=-0.30$, and 
a typical increase of the Ba abundances by 0.13\,dex.

\subsubsection{Astrophysical $\log gf$'s from the Holweger-M\"uller model}
\label{SHMmodel}
As an alternative to the MARCS model, the 
Holweger \& M\"uller (1974) model (HM) was used in a test to derive
astrophysical $\log gf$ values for the lines in the programme.
The $\log gf$ values then became 0.11\,dex higher on the mean compared to the 
ones from a MARCS solar model, with a maximum difference of $+0.20$\,dex.
The stellar abundances decreased by 0.03\,dex to 0.15\,dex, with 
a mean of 0.10\,dex for all species, see Table\,\ref{Tabundchange}. 

Although we have chosen a strictly differential approach in this work, 
with a solar model from the same grid as the model atmospheres representing 
the stars, this choice is not obvious, see sect.\,\ref{Soscillatorstrengths}.
The HM model is known to represent solar limb-darkening and fluxes quite well,
and in fact more successfully than theoretical flux-constant 1D models 
(cf. Blackwell et al. 1995, Gehren et al. 2001,
Edvardsson et al., in preparation).
Thus, there are reasons to believe that the $gf$ values derived from that model
might be better in absolute terms. The cancellation argument, founded on the 
expectation that systematic errors in the stellar model atmospheres act in 
similar ways in the solar analysis, is of uncertain weight. Thus, systematic 
errors in abundances [X/H] of about 0.15\,dex are possible. The corresponding 
errors in abundances relative to iron are smaller; [X/Fe] are on the mean
0.05\,dex higher when derived with the HM model (oxygen not included).

\subsubsection{$\log gf$'s from VALD}
The analysis was also alternatively performed using oscillator strengths taken
from the VALD database.
The $\log gf$ values from VALD were on average only 0.04\,dex lower compared
to the ones used in the programme, but the maximum difference
(us-VALD) was as large as 2.15\,dex (the Fe\,{\sc i} line 7780.552\,\AA) and the
minimum $-0.60$\,dex.
The difference (us-VALD) for neutral iron alone was on the mean $-0.48$\,dex,
with a spread as large as 0.99\,dex.
This had of course a large impact on the abundances derived, especially for
species measured with only one line.
A larger spread and sometimes a different form of the abundance trends in the
diagrams could be seen.
We believe that although the use of astrophysical $\log gf$ values has
drawbacks - they depend on the solar model atmosphere, the damping
and errors introduced in the measurement of equivalent widths -- they are still
to be preferred in a differential analysis.

\subsection{Binaries}
If a star is a binary the absolute magnitude and radius determination will be
affected and the surface gravity, $\log g$, will be underestimated.
The error in the $\log g$ may amount to 0.3\,dex. Also, there may be some 
error in the adopted effective temperature as well as in the metallicity.
None of our spectra show line doubling, which would be an obvious reason
for exclusion from the sample.
We estimate that the abundances for only a few stars should be
significantly affected with abundance errors in excess of 0.05\,dex as a
result of binarity.

\subsection{Errors in the stellar space velocities}
The uncertainties in the total space velocities,
$V_{\rm tot}=\sqrt{U^2+V^2+W^2}$, are estimated to be less than 20\,km/s for
all but ten stars
which all have $V$ velocities lagging behind the LSR by more than 90\,km/s,
$V_{\rm tot}$ larger than 160\,km/s and metallicities less than [Fe/H]$\,=-1.2$.
Another seven stars have uncertainties between 10 and 20\,km/s.
The major source of these errors is uncertain parallax values, while the radial
velocity errors are negligible.

\subsection{Total errors}
It is a difficult task to estimate the total errors in abundance
analyses, since estimates of different types of errors have
to be made and added into simple numbers. From Table\,\ref{Tabundchange} we see
that the errors in effective temperature are significant for abundances that
were derived from atomic lines of elements that are strongly ionized, i.e.
Fe, Na, Mg, Al, Si, Ca, Ti, V, Cr, and Ni. For all these elements the
errors caused by uncertainties in stellar fundamental parameters are 
dominated by the effective-temperature uncertainty, and amount to typically
0.05-0.10\,dex.
In addition to this, errors in $gf$ values as reflected by differences that
result from using different solar models may contribute systematic errors of
typically 0.1\,dex in absolute abundances.
To this comes the fact that departures from LTE and
convectively generated thermal inhomogeneities may well add systematic
errors of typically 0.1\,dex or somewhat more
(Asplund et al. 2004, Gehren et al. 2004).
The errors in abundance ratios, such as 
[Ca/Fe] are hopefully significantly smaller. For the abundance of Sc, 
the gravity uncertainty is also significant and may amount to 0.06\,dex or so, 
but a more significant uncertainty for this element is that its abundance
is based just on one spectral line and thus measurement errors or blends may be
significant.
Also the Ba abundance is based on a single and even blended line with an
uncertain $gf$ value; in this case the uncertainty in microturbulence parameter
is also significant, and the Ba abundance may therefore well be wrong by
0.3\,dex.

A special case is oxygen. Here, effective-temperature errors and
errors in the gravity may possibly add up to about 0.15\,dex.
Since the temperature scale may be systematically in error, this may lead to
systematically erroneous abundances by, say, 0.1\,dex.
In addition to that, there may be errors caused by oversimplified model
atmospheres, notably the assumption of plane-parallel stratification.
We note that Garc\'ia P\'erez et al. (2005) find systematic
deviations for metal-poor subdwarfs between abundances from O\,{\sc i} triplet
lines and [O\,{\sc i}] lines of typically 0.2\,dex and ascribe these differences
to errors in the abundances derived from the triplet lines.   

\section{Comparisons with previous results}
\label{Sdiscussion}
In Fig.\,\ref{Fresults} we display our results together with
abundances for disk stars analysed with very similar data and methods.
A general observation is that the transitions between the disk and halo
population data are smooth and continuous for all elements.

We have compared our results with those of a number of recent studies which
have some stars in common with ours. The results of these comparisons are
displayed in Table\,\ref{Tabundcomp}.
\begin{table}
\caption[]{\label{Tabundcomp}
Comparison between our abundance results and those of others for stars in common.
The numbers of stars, the mean differences and standard deviations
are given.  Sources: B: Burris et al. (2000), BG: Baum\"uller \& Gehren (1997),
F: Fuhrmann (1998), Ge: Gehren et al. (2004), Gr: Gratton et al. (2003a),
J: Jehin et al. (1999), MG: Mashonkina \& Gehren (2001), McW: McWilliam (1998),
N: Nissen et al. (2002), NS: Nissen \& Schuster (1997).
}
\begin{flushleft} 
\begin{tabular*}{8.8cm}{llrrr}
Element~~~~~~~ & Source~~ & ~~~~~N\;\; & $\Delta$[X/Y](us-Source) &~~~~~\\
\hline
[Fe/H] Fe\,{\sc i}  & Gr     & ~~~~~17\;\;   & $ 0.02 \pm 0.13$ ~~~~~\\
                    & Gr     & ~~~~~15\hidewidth$^1$ & $-0.02 \pm 0.08$ ~~~~~\\
                    & J      & ~~~~~13\;\; & $-0.03 \pm 0.04$ ~~~~~\\
                    & N      & ~~~~~ 8\;\; & $ 0.01 \pm 0.07$ ~~~~~\\

[Fe/H] Fe\,{\sc ii} & N      & ~~~~~ 6\;\; & $ 0.00 \pm 0.04$ ~~~~~\\

[O/Fe]              & B$^2$  & ~~~~~ 5\;\; & $ 0.05 \pm 0.19$ ~~~~~\\
                    & B$^3$  & ~~~~~ 5\;\; & $-0.02 \pm 0.15$ ~~~~~\\
                    & Gr     & ~~~~~15\;\; & $ 0.00 \pm 0.12$ ~~~~~\\
	            & N$^4$  & ~~~~~ 2\;\; & $ 0.01 \pm 0.02$ ~~~~~\\
	            & NS     & ~~~~~ 3\;\; & $ 0.26 \pm 0.06$ ~~~~~\\

[Na/Fe]             & Gr     & ~~~~~16\hidewidth$^5$ & $ 0.03 \pm 0.11$ ~~~~~\\
                    & Ge$^6$ & ~~~~~ 4\;\; & $-0.10 \pm 0.06$ ~~~~~\\
                    & NS     & ~~~~~ 3\;\; & $ 0.03 \pm 0.03$ ~~~~~\\

[Al/Fe]             & BG     & ~~~~~ 3\;\; & $-0.14 \pm 0.13$ ~~~~~\\

[Mg/Fe]             & Gr     & ~~~~~12\hidewidth$^7$ & $-0.06 \pm 0.07$ ~~~~~\\
                    & Ge$^6$ & ~~~~~ 4\;\; & $-0.03 \pm 0.12$ ~~~~~\\
                    & F      & ~~~~~ 2\;\; & $-0.05 \pm 0.06$ ~~~~~\\
	            & J      & ~~~~~12\;\; & $ 0.05 \pm 0.06$ ~~~~~\\
                    & NS     & ~~~~~ 3\;\; & $ 0.05 \pm 0.04$ ~~~~~\\

[Si/Fe]             & Gr     & ~~~~~14\hidewidth$^8$ & $-0.04 \pm 0.06$ ~~~~~\\
                    & NS     & ~~~~~ 3\;\; & $ 0.05 \pm 0.01$ ~~~~~\\

[Ca/Fe]             & Gr     & ~~~~~15\hidewidth$^9$ & $-0.04 \pm 0.04$ ~~~~~\\
                    & J      & ~~~~~13\;\; & $ 0.04 \pm 0.04$ ~~~~~\\
                    & NS     & ~~~~~ 3\;\; & $ 0.04 \pm 0.01$ ~~~~~\\
 
[Sc/Fe]             & Gr     & ~~~~~ 4\;\; & $ 0.11 \pm 0.05$ ~~~~~\\

[Ti/Fe]             & Gr     & ~~~~~ 6\;\; & $ 0.05 \pm 0.05$ ~~~~~\\
                    & J      & ~~~~~ 4\;\; & $ 0.05 \pm 0.04$ ~~~~~\\
                    & NS     & ~~~~~ 2\;\; & $ 0.10 \pm 0.02$ ~~~~~\\

[V/Fe]              & Gr     & ~~~~~ 8\;\; & $ 0.13 \pm 0.13$ ~~~~~\\
                    & J      & ~~~~~ 9\;\; & $ 0.19 \pm 0.06$ ~~~~~\\

[Cr/Fe]             & Gr     & ~~~~~ 5\;\; & $ 0.15 \pm 0.11$ ~~~~~\\
                    & J      & ~~~~~ 8\;\; & $ 0.09 \pm 0.09$ ~~~~~\\
                    & NS     & ~~~~~ 3\;\; & $ 0.07 \pm 0.12$ ~~~~~\\

[Ni/Fe]             & Gr     & ~~~~~14\;\; & $ 0.01 \pm 0.08$ ~~~~~\\
                    & J      & ~~~~~13\;\; & $ 0.09 \pm 0.06$ ~~~~~\\
                    & NS     & ~~~~~ 3\;\; & $ 0.01 \pm 0.03$ ~~~~~\\

[Ba/Fe]             & B      & ~~~~~ 2\;\; & $-0.25 \pm 0.37$ ~~~~~\\
                    & J      & ~~~~~12\;\; & $-0.16 \pm 0.15$ ~~~~~\\
                    & MG     & ~~~~~ 2\;\; & $-0.20 \pm 0.11$ ~~~~~\\
                    & McW    & ~~~~~ 2\;\; & $ 0.14 \pm 0.18$ ~~~~~\\
                    & NS     & ~~~~~ 3\;\; & $-0.05 \pm 0.08$ ~~~~~\\
\hline
\multicolumn{4}{l}{$^1$ \object{HD\,140283} ($\Delta=0.35$), \object{HD\,84937}
                   ($\Delta=0.22$\,dex) excluded} \\
\multicolumn{4}{l}{$^2$ $T_{\rm eff}$ scale according to King (1993)} \\
\multicolumn{4}{l}{$^3$ $T_{\rm eff}$ scale according to Carney (1983)} \\
\multicolumn{4}{l}{$^4$ [O/H] (NLTE) from IR triplet lines} \\
\multicolumn{4}{l}{$^5$ \object{HD\,84937} ($\Delta= 0.53$) excluded} \\
\multicolumn{4}{l}{$^6$ LTE abundances} \\
\multicolumn{4}{l}{$^7$ \object{HD\,94028}, 132475, 140284
                   ($\Delta = 0.19-0.3$) excluded} \\
\multicolumn{4}{l}{$^8$ \object{HD\,116064} ($\Delta=-0.28$) excluded} \\
\multicolumn{4}{l}{$^9$ \object{HD\,84937} ($\Delta=-0.17$), \object{HD\,140283}
                   ($\Delta=-0.38$) excluded} \\

\end{tabular*}
\end{flushleft}
\end{table}
Although these studies all share a number of
basic features (basically similar model assumptions), they deviate
partly in terms of details, such as temperature calibrations, $gf$ values,
and model atmospheres. We therefore note with satisfaction that the mutual
agreement for stars in common is good. We shall now comment more specifically
on our results as compared with those of others. 

\subsection{Iron}
Our abundances of iron were based on Fe\,{\sc i} lines, in spite of the fact
that these may be subject to NLTE effects, due to over-ionisation 
(e.g. Saxner \& Hammarb\"ack, 1985, Th\'evenin \& Idiart, 1999) or departures
from LTE excitation (Magain \& Zhao, 1996).
The reason for this choice was simply that just one Fe\,{\sc ii} line was
observed.
This line, however, could be measured in all but six stars.
The resulting iron abundances were close to those
measured from the Fe\,{\sc i} lines, the mean difference being $+0.03$\,dex,
with a standard deviation of 0.07\,dex.
This result, of a practically null difference, agrees with the finding by
Gratton et al. (2003a) and Jehin et al. (1999) and departs significantly from
the result of the 
calculations by Th\'evenin \& Idiart who find effects ranging from 0.15\,dex
to 0.3\,dex for the range in metallicity of our stars. We note, however, that
our result depends on the $gf$ values adopted -- if we instead of the solar
MARCS model use the HM model for the Sun, we derive typical over-ionization
effects in [Fe\,{\sc ii}/Fe\,{\sc i}] of 0.1\,dex for the programme stars. 

As is clear from Sect.\,\ref{Smetalerror} above, we find little evidence of
departures from the Boltzmann excitation equilibrium, in contrast with the
empirical results of Magain \& Zhao (1996).
The latter authors have adopted $gf$ values for the lower excitation iron lines
from Blackwell et al. (1995); for the higher excitation lines, however, 
they derive $gf$ values from fitting observed solar-disk centre equivalent
widths to the HM model, adopting a relatively
high solar abundance ($\log \epsilon_\odot({\rm Fe})=7.68$).
As these authors note, a systematical lowering of the solar 
abundance of iron to a value around 7.5, as we have chosen, would nearly 
cancel their excitation-equilibrium effect for the metal-poor stars, while 
consistency with a standard LTE analysis with the HM model would not be 
obtained. 

\begin{figure}
\centering
\resizebox{\hsize}{!}{\includegraphics*{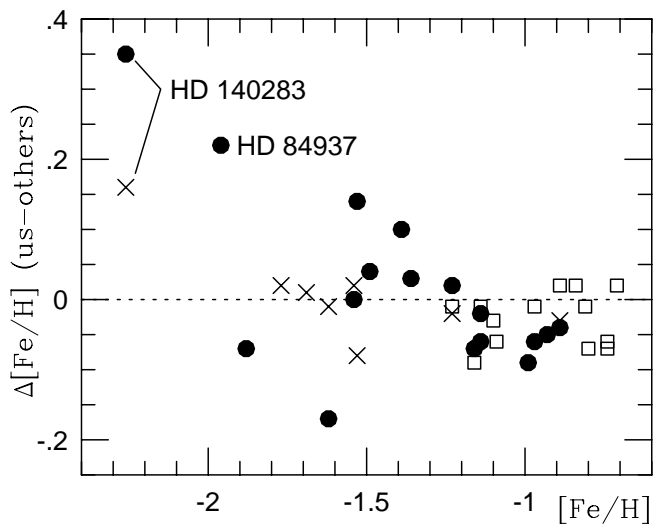}}
\caption[]{\label{Fironcomp}
Comparison of iron abundances derived from Fe\,{\sc i} lines with those of
Jehin et al. 1999 (squares) and Gratton et al. 2003a (dots).
The differences relative to abundances from lines of Fe\,{\sc ii} of
Nissen et al. 2002 are shown by crosses.}
\end{figure}

As may be seen from Table\,\ref{Tabundcomp} and Fig.\,\ref{Fironcomp},
we find an overall agreement in iron abundances
with Nissen et al. (2002), Gratton et al. (2003a) and Jehin et al. (1999). 
The abundances of Nissen et al. were based on about 10 Fe\,{\sc ii} lines
while those of
Gratton et al. are, at least to a great extent, based on Fe\,{\sc i} lines.
The star with the largest departure by far in the comparison with 
Gratton et al. is \object{HD\,140283} for which we 
find a considerably higher effective temperature which explains parts of 
the difference. The second largest departure occurs
for \object{HD\,84937}, which is the hottest star in our sample. 

Jehin et al. (1999) used Fe\,{\sc i} lines to derive [Fe/H] abundances
differentially relative to \object{HD\,76932}, for which a metallicity of
[Fe/H]\,$=-0.91$ was derived from laboratory $gf$ values
(our value for this star is $-0.89$).
They also derived iron abundances from Fe\,{\sc ii} lines; for the
13 stars we have in common
their mean value of [Fe/H] from Fe\,{\sc ii} is $-0.02$\,dex lower
than that obtained from the Fe\,{\sc i} lines, i.e. no signs of
over-ionisation which is also consistent with our result. 

We also have 13 stars (to some extent the same stars as with the Jehin et al.
study) in common with the earlier study of Zhao \& Magain (1991).
Our values of [Fe/H] depart systematically from those of Zhao \& Magain by,
on the mean, as much as $+0.19$\,dex.
This difference may be ascribed to the choice of $gf$ values and 
solar iron abundance of Zhao \& Magain; they use the HM model to derive $gf$ 
values for the higher excitation lines and adopt a solar iron abundance, 
$\log \epsilon({\rm Fe})$, of 7.66.
The differences between the results of the studies by
Zhao \& Magain (1991) and Jehin et al. (1999) reflect a basic uncertainty
in the solar analysis of iron.

\subsection{Oxygen}
Our oxygen abundances, as plotted in the [O/Fe] vs. [Fe/H] diagram depart to
some extent from the studies of Boesgaard et al. (1999) and
Israelian et al. (1998). 
These latter authors estimated the oxygen abundances 
in metal-poor stars from OH lines in the ultraviolet region
and found a tendency for a systematically increasing [O/Fe] with 
decreasing [Fe/H], with [O/Fe] values approaching $+1.0$ for the most
metal-poor stars with [Fe/H] around $-3$.
Here, we shall not comment further on the OH oxygen-abundance determinations
which may be severely plagued by uncertainties in UV opacities, convective
inhomogeneities etc. (Asplund et al. 1999, Nissen et al. 2002,
Asplund \& Garc\'ia P\'erez 2001).
Boesgaard et al. also made a literature study of the abundances 
from the oxygen triplet of their programme stars and applied two different
temperature scales, due to Carney (1983) and King (1993), respectively.
There are some differences in our analyses of triplet lines
and those of Boesgaard et al. A major difference is that we
correct for NLTE effects which Boesgaard et al. do not.
Boesgaard et al. used a higher oxygen abundance for the Sun than 
we do, but this is compensated for by our higher $gf$ values.
They also use Kurucz models while we use MARCS models.
Boesgaard et al. investigated the difference between the two 
stellar model grids and concluded that 
the abundances are $0.07\pm0.03$\,dex higher when the Kurucz models
are used than for MARCS models, which agrees with our own estimates. 
Also, there are differences in stellar parameters.
For the five stars in common between the two studies our
temperatures are on average $47\pm73$\,K lower compared with the King
temperature scale, and $85\pm67$\,K higher compared with the Carney scale. 
Our gravity parameters and metallicities are very similar, as are the
observed equivalent widths.

Here we first compare the result of the LTE analyses.
For the 5 stars in common between us and Boesgaard et al. (B), 
the mean difference [O/H]$_{\rm LTE}$(us--B) is $-0.02$ with a scatter
(s.d.) of $\pm0.17$\,dex for the King temperature scale,
and $-0.09\pm0.13$\,dex for the Carney scale. 
The different stellar models, in addition to the different temperature
scales, are the most important reason for the differences in LTE abundances
from the oxygen triplet lines. 

When we compare our [O/H] abundances from the triplet lines, as 
corrected for NLTE effects (for the Sun as well as for
the stars), they again agree well on average with
those of Boesgaard et al. (Table\,\ref{Tabundcomp}).
\begin{figure}
\centering
\resizebox{\hsize}{!}{\includegraphics*{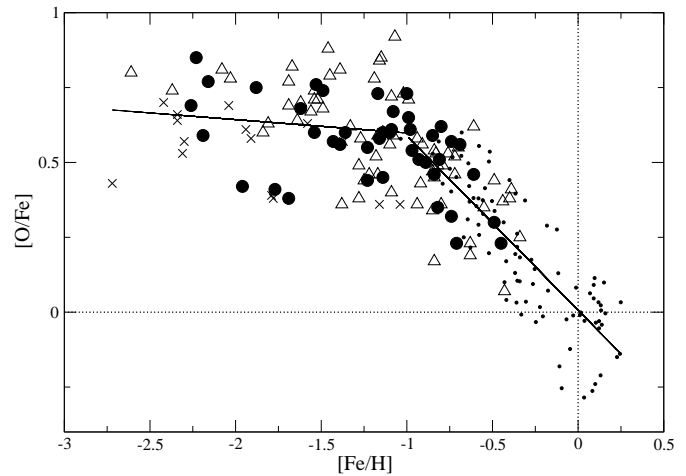}}
\caption[]{\label{FseveralprojectsOFe-FeH}
Our oxygen abundances relative to iron, based on the strength of the IR
7774\,{\AA} triplet, and with corrections for NLTE applied (filled circles).
The values derived with a very similar method by Nissen et al. (2002) are
represented by crosses, and those of Gratton et al. (2003a) by triangles.
The disk-star data of Edvardsson et al. (1993) (which were empirically
normalized to NLTE via the [O\,{\sc i}] 6300\,\AA\ line) is shown as small dots.
The lines are the linear least-squares fits of Eq.\,\ref{Eoxygenfitall}.
}
\end{figure}

Gratton et al. (2003a) have also published oxygen abundances for Pop II stars
based on the triplet lines. 
Corrections for NLTE effects have been calculated. These authors claim that 
the resulting corrections are consistent with those calculated by
Nissen et al. (2002) and so they ought also to be consistent with our
corrections.
We find an overall fair agreement with the results of Gratton et al.
(Fig.\,\ref{FseveralprojectsOFe-FeH} and Table\,\ref{Tabundcomp}).
As will be commented  on below, the scatter (0.12\,dex) for the stars in common
may be due to errors in equivalent widths.  
We note, however, in passing that Gratton et al. 
derive an oxygen abundance $0.12\pm0.04$\,dex higher from the triplet lines than
they obtain from the forbidden 6300\,{\AA} line, an effect which they ascribe 
to reddening and binarity which should affect the triplet lines more than the 
forbidden one. 
 
Nissen et al. (2002) have derived oxygen abundances from the triplet lines and 
alternatively from the forbidden 6300\,{\AA} line.
NLTE corrections for the triplet lines were applied  which are fully consistent
with ours.
Nissen \& Schuster (1997) obtained oxygen abundances from the triplet lines in
an LTE analysis, however, differentially versus two Pop\,II dwarfs with O 
abundances derived on a scale based on the forbidden lines. Since their range 
in parameters was rather limited, the authors argue that most of the NLTE 
effects of the resulting oxygen abundances from the triplet lines should be 
compensated for.
While we find a good agreement for the stars in common with
Nissen et al. (2002), the comparison with Nissen \& Schuster (1997) shows a mean
difference of 0.26\,dex and a scatter of 0.06 for the 3 stars in common. 
When comparing [O/Fe] values by Nissen \& Schuster with those 
derived from the forbidden line of Nissen et al. (2002) we note that the 
latter values are systematically higher by 0.13\,dex.
We conclude that the [O/Fe] scale of Nissen \& Schuster (1997) seems to be
somewhat lower than that of Nissen et al. as well as for the present study. 
A reason for this may be that the assumed [O/Fe]\,$=0.30$ for the standard stars
of Nissen \& Schuster departs from that of Nissen et al.; for one of these
stars (\object{HD\,76932}) Nissen et al. (2002) give the value [O/Fe]\,$=0.44$
while our value is 0.38. 

In Fig.\,\ref{FseveralprojectsOFe-FeH} we have plotted our values of [O/Fe] vs.
[Fe/H] and compared with all stars with triplet abundances in the samples of 
Gratton et al. (2003a) and of Nissen et al. (2002).
As expected from the agreement for stars in common we find a general good
agreement.
Our total equivalent widths for the triplet lines show a scatter relative to the
widths of Gratton et al. (2003a) for the stars in common by about 20\%,
leading to an abundance scatter of about 0.12\,dex.
It is unlikely, however, that all the scatter in
Fig.\,\ref{FseveralprojectsOFe-FeH} at a given [Fe/H] is due to errors in
equivalent widths. The oxygen triplet lines are very sensitive to errors in 
effective temperature and gravity. 
Errors assumed to be about 100\,K in $T_{\rm eff}$ and 0.15\,dex in $\log g$
lead to errors in [O/H] of typically 0.1\,dex.
Although there may be errors in model atmospheres and NLTE corrections that
vary systematically with the $T_{\rm eff}$, $\log g$ and [Fe/H] and cause 
systematic errors in the oxygen abundances, it is questionable whether these are
able to increase the scatter to that observed in
Fig.\,\ref{FseveralprojectsOFe-FeH}.
In particular we note three stars, \object{HD\,84937}, \object{HD\,122196} and
\object{HD\,160617} with [Fe/H]\,$\sim -2$ that seem to have lower 
[O/Fe] values than other stars at these metallicities.
Of these stars \object{HD\,84937} has the highest positive U velocity of all
stars in the sample.  \object{HD\,160617} was also observed by
Nissen et al. (2002) and found to have a comparatively low oxygen abundance. 

We have also plotted points representing the disk stars analysed by
Edvardsson et al. (1993) (Fig.\,\ref{FseveralprojectsOFe-FeH}).
What tendency with [Fe/H] can be traced in this diagram? 
It is seen that the slope for the disk is not clearly broken by a constant
plateau with constant [O/Fe] at [Fe/H]\,$\sim -1$, as suggested by e.g.
Barbuy (1988).
Our data instead support a tendency for a continued, although much less steep
gradient towards lower metallicities than that suggested by
Israelian et al. (1998) and Boesgaard et al. (1999) on the basis of measurements
of OH lines in the ultraviolet.
From fits to our data (filled circles) and those of Edvardsson et al.
(small dots) we obtain the following mean linear relations for the regions
below and above [Fe/H]\,$=-1.0$:
\begin{align}
\label{Eoxygenfitus}
-2.75 < [{\rm Fe/H}] &  < -1.0:   \nonumber \\   
[{\rm O/Fe}]&=0.52-0.06\,[{\rm Fe/H}], ~ \sigma_{\rm std}=0.12 \nonumber \\ 
-1.0 < [{\rm Fe/H}] &  < 0.25:   \nonumber \\ 
[{\rm O/Fe}]&=0.00-0.60\,[{\rm Fe/H}], ~ \sigma_{\rm std}=0.12\,,
\end{align}
where $\sigma_{\rm std}$ denotes the standard deviation of [O/Fe] from the line.
The uncertainties in the slope coefficients for the two lines are,
respectively, 0.06 and 0.03.
Including also the data from Nissen et al. (2002) (diamonds) and
Gratton et al. (2003a) (triangles) the fits remain quite similar:
\begin{align}
\label{Eoxygenfitall}
-2.75 < [{\rm Fe/H}]& < -1.0:  \nonumber \\
[{\rm O/Fe}]&=0.55-0.04\,[{\rm Fe/H}], ~  \sigma_{\rm std}=0.14 \nonumber \\
-1.0 < [{\rm Fe/H}]&< 0.25:  \nonumber \\
[{\rm O/Fe}]&=0.01-0.58\,[{\rm Fe/H}], ~ \sigma_{\rm std}=0.12\,.
\end{align}
Here the uncertainties in the slope coefficients are 0.04 and 0.03,
respectively.
We note that our results are consistent with e.g. Barbuy (1988) in that no
significant systematic variation of [O/Fe] with [Fe/H] among halo stars is
found.

Finally, we note that Garc\'ia P\'erez et al. (2005) have recently determined
oxygen abundances for Pop\,II subgiants, for which three different
oxygen-abundance criteria have been possible to measure: the IR triplet
O\,{\sc i} lines, the forbidden 6300.3\,\AA\ line and the OH UV molecular lines.
For the most metal-poor subgiants there seems to be a tendency for the triplet
lines to indicate higher O abundances than the other criteria, which is contrary
to the effect found by Boesgaard et al. (1999) for hotter stars.
This effect is not fully understood, but should be taken as a warning that our
oxygen abundances may still suffer from systematic errors.

\subsection{Sodium}
\label{Ssodium}
\begin{figure}
\centering
\resizebox{\hsize}{!}{\includegraphics*{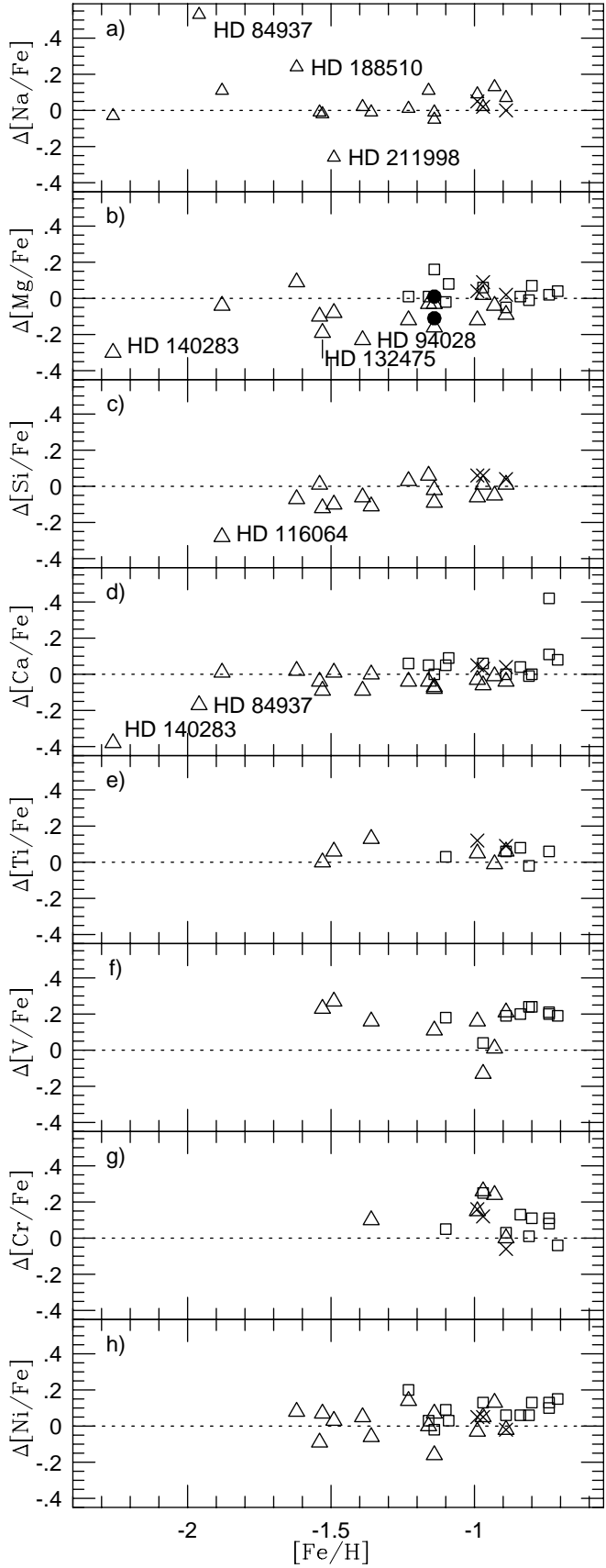}}
\caption[]{\label{Fabcomp}
Comparison of our derived values of [X/Fe] with those of
Nissen \& Schuster 1997 (crosses), Fuhrmann 1998, (filled circles),
Jehin et al. 1999 (squares) and Gratton et al. 2003a (triangles).
$\Delta$[X/Fe]\,$=$\,[X/Fe](us)$-$[X/Fe](others).}
\end{figure}
Our resulting [Na/Fe] are compared with those of 
Gratton et al. (2003) and Nissen \& Schuster (1997) in Table\,\ref{Tabundcomp}
and Fig.\,\ref{Fabcomp}\,a. Except for three outliers there is a good agreement.
\object{HD\,84937} is one of the 3 stars in the full sample (150 stars) of 
Gratton et al. for which the authors give a value of [Na/Fe] below $-0.40$, and
we recommend a closer inspection of the spectra of this star. 

Our [Na/Fe] values are typically about $+0.2$ for stars around [Fe/H]\,$=-0.8$,
a value which agrees with the results for the disk by Edvardsson et al. (1993)
(see Fig.\,\ref{Fresults}b) and which may be typical both for thin and thick
disk stars (Bensby et al. 2003).
At lower metallicities, however, [Na/Fe] decreases to 0.0 or close to that,
which agrees with the result of McWilliam et al. (1995) for even more metal-poor
stars.
Baum\"uller, Butler \& Gehren (1998) have
studied the NLTE effects on Na abundances for metal-poor stars,
and found abundance corrections of typically $-0.07$\,dex for
abundances derived from the 5682/88\,\AA\ doublet and about half of that
from the 6154/60\,\AA\ doublet for stars in our parameter range.
The effects on the corresponding line in the solar spectrum
were found to be marginal. The effects on our most metal-poor stars
may be somewhat larger, though probably not very important.
Gehren et al. (2004) have recently recalculated their non-LTE corrections for 
abundances of Na, Al and Mg for metal-poor stars, with better representations of
atoms and of UV fluxes.
We note that the [Na/Fe] values for our most metal-poor stars are significantly
higher by about 0.2\,dex than the NLTE values in Gehren et al. (2004) for their
most metal-poor stars, which certainly is a result of their correction for
NLTE.
For the four stars we have in common, the mean difference us -- Gehren et al.
is thus $+0.17$\,dex; if the LTE values of Gehren et al. are compared with ours,
the mean difference is instead $-0.10$\,dex according to
Table\,\ref{Tabundcomp}.

\subsection{Aluminium}
For disk metallicities ([Fe/H]\,$\ga -1.0$) our data smoothly join the trend of
Edvardsson et al. (1993) in Fig.\,\ref{Fresults}d.
At lower metallicities, however,
the trend in [Al/Fe] with [Fe/H] found by us is not consistent with the trends 
found repeatedly by others for field stars (e.g. Aller \& Greenstein 1960, 
Arpigny \& Magain 1983, Spite \& Spite 1980, Gratton \& Sneden 1988, and
Shetrone 1996), who find systematically decreasing [Al/Fe] with decreasing 
metallicity with values of [Al/Fe] close to 0.0 at [Fe/H]\,$\sim -1.5$.
The situation was summarised by McWilliam (1997). The tendency found by these 
authors is generally interpreted as a result of the metallicity dependent Al 
yields predicted for carbon burning by Arnett (1971), although the similar 
prediction that Na should also decrease with metallicity has not been verified.
Even though a tendency for decreasing Al may be traced in
Fig.\,\ref{Fresults}
for the lowest metallicities, our results depart so much from the previous
results that it is important to trace the reason for the differences.
The authors mentioned 
have no stars directly overlapping with those of our sample for which Al has 
been determined. While most of the previous authors have used the ultraviolet 
resonance lines of Al, we have used two infrared lines with 4\,eV excitation 
energies (an exception is the work by Shetrone (1996), who, however, analysed 
Pop\,II giant stars). Baum\"uller \& Gehren (1997) have performed 
statistical-equilibrium calculations for Al in Pop\,II stars. They find very 
severe effects on the resonance lines, and suggest abundance revisions upwards 
for determinations from these lines of 0.4 -- 0.6\,dex for solar-type Pop\,II 
stars. For the high-excitation lines the corrections are much smaller, on the 
order of 0.1\,dex. With such corrections applied, our values will be in fair 
consistency with corrected values from the literature. The calculations of 
Baum\"uller \& Gehren still suffer from uncertainties in collision cross 
sections, not the least for collisions with hydrogen atoms. In fact, for the
three stars in common with them, we find a mean difference of -0.14 
(cf. Table\,\ref{Tabundcomp}) suggesting that their NLTE effects
are somewhat overestimated; for
these stars, the BG determinations are based on the near-UV resonance lines.
In comparing our results with those of Gehren et al. (2004) we note that our few
stars with $-1.6 \leq$\,[Fe/H]\,$\leq -1.0$ tend to lie above their
few stars by $\Delta\,$[Al/Fe]\,$\sim 0.2$\,dex.
Our results support, 
however, their main conclusion on the significance of departures from LTE, 
although problems may remain in fitting the results from models of galactic 
nucleosynthesis by Timmes et al. (1995). 

\subsection{Magnesium, silicon and calcium}
\label{MgSiCa}
In general we find a good agreement with Gratton et al. (2003a),
Jehin et al. (1999) and Nissen \& Schuster (1997) as regards [Mg/Fe], [Si/Fe]
and [Ca/Fe] for the common stars (cf. Table\,\ref{Tabundcomp}.
Fig.\,\ref{Fabcomp}\ b, c and d). For three outliers, however, Gratton et al.
give [Mg/Fe] values in excess of ours by 0.19 -- 0.3\,dex.
These differences basically reflect differences in [Fe/H]; if we
compare [Mg/H] values instead the 3 stars agree well with differences ranging
from $-0.13$ to $+0.05$, and a mean of $-0.06$\,dex. 
The agreement in Mg for stars in common with Nissen \& Schuster (1997) 
and Jehin et al. (1999) is good, although their abundances are based on 
measurements of one single line.
Unfortunately, we have only 2 stars in common with the detailed study of
Fuhrmann (1998).

Gehren et al. (2004) have found relatively limited but significant NLTE
corrections upwards for their Mg abundances of nearby metal-poor stars.
Typically, the increase in [Mg/Fe] is 0.1 -- 0.2\,dex for their most metal-poor
stars.
We have four stars in common with their study, and find a mean difference
$\Delta$\,[Mg/Fe](us - Gehren) of $-0.16$\,dex; adopting the LTE values of
Gehren et al. we would have obtained $-0.03$\,dex instead. 

We note that the NLTE corrections calculated by Gehren et al. are metallicity
dependent -- from their Table\,3 we derive
$\Delta$(NLTE-LTE)\,$\sim -0.07\cdot$\,[Fe/H].
If this correction were applied to the points in Fig.\,\ref{Fresults}, 
a slowly sloping trend in [Mg/Fe] for stars with
[Fe/H]\,$\leq -1.0$ seems to emerge.
Such a trend may also be present for the smaller numbers of stars in
Gehren et al. (2004), see their Fig.\,6, middle right panel.

For the Si abundances of stars in common with Gratton et al. (2003a) we also
find a limited mean deviation, and a relatively small scatter, provided that 
the single outlier is excluded. 
For stars in common with Nissen \& Schuster (1997) we also find a good
agreement. 

Similarly, for the stars with Ca abundances in common with
Gratton et al. (2003a) we find a good agreement, if two outliers 
with significantly smaller [Ca/Fe] values in our
study are excluded. Again, both these 
stars deviate due to the differences in [Fe/H], not in [Ca/H] for which 
they only deviate by 0.05\,dex or less. If all common stars are included, 
the mean difference in [Ca/H] is $-0.03$ with a scatter of $0.07$.
For stars in common with
Jehin et al. (1999) and with Nissen \& Schuster (1997) we also find only small
deviations.

In spite of the relatively marginal differences in abundances between the
different studies, we observe some suspicious regularities.
The Gratton et al. values of [Mg/Fe], [Si/Fe] and [Ca/Fe] all tend to be higher
than those of ours by, at the most, $0.06$\,dex , while the Nissen \& Schuster
values as well as the Jehin et al. values are below those of ours by
about the same amount.
Some of these differences may be due to different choices of
stellar fundamental parameters or model atmospheres, or of spectral lines.
It seems clear that even if one disregards errors due to 1D LTE model
atmospheres with mixing-length convection, realistic systematic errors in
abundance analyses for Mg and Si in Pop\,II stars are of this order of
magnitude, while they may be smaller for the spectroscopically more ideal
element Ca.

The trends and scatter for [Mg, Si, Ca/Fe] will be discussed in
Sect.\,\ref{Scorrelations}.

\subsection{Scandium, titanium, vanadium and chromium}
The elements Sc, V and Cr are represented by just one spectral feature each,
and Ti is represented by 2 lines in our study. Also, the number of spectra 
in which the features could be measured was limited. 

As seen in Table\,\ref{Tresults}, abundances from the single Sc\,{\sc ii} line
were only obtained for 9 stars. 
The results indicate an iron-like behaviour (i.e., with [Sc/Fe]\,$\sim 0.0$
except for the most metal-poor star \object{HD\,196944} for which a lower Sc
abundance is suggested, Fig.\,\ref{Fresults}).
The slowly rising trend in [Sc/Fe] with decreasing metallicity
that may be traced in the study of disk stars by
Reddy et al. (2003, their Fig.\,10) does not seem to continue for Pop\,II stars.
This result departs from the early result of Magain (1989, his Fig.\,9) but
agrees with the more recent compilation of
Norris, Ryan \& Beers (2001, their Fig.\,8).
The general behaviour also agrees with e.g. Cayrel et al. (2004) 
and Barklem et al. (2004), extending to the most metal-poor stars with a
remarkably small scatter. However, we note that both
Cayrel et al. and Barklem et al. find a mean [Sc/Fe] somewhat above 0.0,
as in fact also Norris, Ryan \& Beers suggested for the most metal-poor stars.
Our result agrees with the general result of Gratton et al. (2003a),
although a direct comparison for the 4 stars in common shows that our values are
higher by 0.11\,dex on the mean (cf. Table\,\ref{Tabundcomp}).

The abundances derived from our two Ti\,{\sc i} lines indicate an
$\alpha$-element behaviour with positive values of [Ti/Fe] for the metal-poor
stars, as was traced e.g. for disk stars by Edvardsson et al. (1993), and found
for Pop\,II stars in several studies (see the compilation by
Norris, Ryan \& Beers 2001, their Fig.\,5).
For the stars in common with Gratton et al. (2003a) and Jehin et al.(1999)
the Ti abundances agree well while for
the 2 stars in common with Nissen \& Schuster we obtain higher [Ti/Fe]
values by 0.09 and 0.12\,dex in [Fe/H], respectively, Fig.\,\ref{Fabcomp}\,e
and Table\,\ref{Tabundcomp}.

The V abundances obtained by us also suggest systematically positive [V/Fe]
values for most Pop\,II stars, by typically 0.2\,dex. Comparisons with other 
data for common stars
cast, however, severe doubts on this result, see Fig.\,\ref{Fabcomp}\,f and
Table\,\ref{Tabundcomp}.
There is no support from these comparisons that [V/Fe] would be systematically
above 0.0
This would also be unexpected, since V is regarded as an iron-peak
element and should scale with iron.
We also note that Fulbright (2000) found [V/Fe] values close to solar for his 
large 
sample of Pop\,II field stars, most of them dwarfs, as did Johnson (1999) for
field giants and Ivans et al. (2001) for globular-cluster giants. 
As was noted by
Ivans et al. (1999) the relatively low ionization energy of V makes
[V/Fe] sensitive to the effective temperature adopted for the star; however,
as may be seen from Table\,\ref{Tabundchange}, the higher excitation
energy of the line used here reduces the temperature sensitivity so much that an
error of 500\,K or more is needed to reduce our [V/Fe] values to 0.0.
We consider it more probable that our systematically high V
abundances may reflect measurement errors due to the weakness of the single
V\,{\sc i} line, with an equivalent width smaller than 10\,m{\AA} for almost all
stars.

Also our Cr abundances seem from Fig.\,\ref{Fabcomp}\,g and Table\,\ref{Tabundcomp} 
somewhat higher than those derived by others.
Again, we may possibly ascribe this
result to the difficulty in measuring the single line in our spectra. Anyhow, we
have no indication that [Cr/Fe] departs from 0.0 for the Pop\,II stars, and we
can not verify nor disprove the finding by several authors 
(e.g., McWilliam et al. 1995, Ryan et al. 1996, Carretta et al. 2002,
Fulbright 2002, Cayrel et al. 2004 and Barklem et al. 2004) that this abundance
ratio becomes subsolar for the most metal-poor stars.

\subsection{Nickel}
Our Ni abundances are based on a maximum of 7 Ni\,{\sc i} lines, of different 
excitation and different strengths.
As many previous authors have found, we find that
the abundance ratios [Ni/Fe] stay close to 0.0 at a mean, although there
may be a tendency for the scatter in the abundance ratio to increase as
[Fe/H] goes below $-1$. We find a fair agreement with the results of
Gratton et al. (2003a), Jehin et al. (1999) and Nissen \& Schuster (1997) 
(Fig.\,\ref{Fabcomp}\,h, Table\,\ref{Tabundcomp}).
It is noteworthy that the two stars in Fig.\,\ref{Fresults} departing most from
the trends, with [Fe/H]\,$=-1.16$ and $-1.14$, respectively, 
are given similarly low Ni abundances by Jehin et al. (1999).
These stars are \object{HD\,193901} and \object{HD\,194598} and have [Ni/Fe]
close to $-0.20$.  Gratton et al. (2003a) also find [Ni/Fe]\,$=-0.20$ for
\object{HD\,193901}, while they obtain [Ni/Fe]\,$=-0.05$ for
\object{HD\,194598}.
In view of the generally fair agreement, we conclude that the 
low [Ni/Fe] values of \object{HD\,193901} and \object{HD\,194598} seem
relatively well established. 

\subsection{Barium}
\label{Sbarium}
\begin{figure}
\centering
\resizebox{\hsize}{!}{\includegraphics*{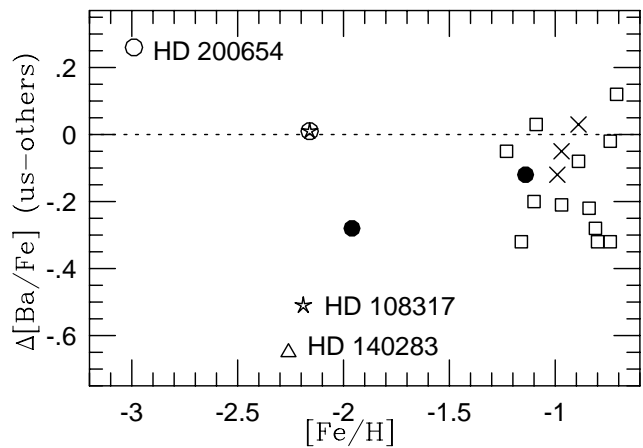}}
\caption[]{\label{Fbacomp}
Comparison of barium abundances with those of
Gratton \& Sneden 1994 (triangle), Nissen \& Schuster 1997 (crosses),
McWilliam 1998 (circles), Jehin et al. 1999 (squares),
Burris et al. 2000 (stars) and Mashonkina \& Gehren 2001 (dots).}
\end{figure}

Our Ba abundances, based on only one blended spectral line with negligible
hyperfine-structure and isotopic wavelength shifts
(cf. Sect.\,\ref{Sanalysingprogram}), show a general behaviour
with [Fe/H] in Fig.\,\ref{Fresults} which is similar to those found by
Nissen \& Schuster (1997) and McWilliam (1998, his Fig. 2b) while they are a
little higher than those of Gratton \& Sneden (1988, 1994) and smaller than
those of Jehin et al. (1999) and Burris et al. (2000). 
Thus, we find a rather small scatter for [Fe/H]\,$>-1.5$ with [Ba/Fe] solar or
slightly sub-solar and a single outlier with high barium abundance, while the
mean [Ba/Fe] may decrease, and the scatter increase, for lower metallicities.

When comparing our abundances with those derived by others for common stars
we find, however, systematic deviations.
These comparisons are summarised in Fig.\,\ref{Fbacomp} and Table\,\ref{Tabundcomp}.
The basic reason for most of these discrepancies seems to be that the
Ba\,{\sc ii} 6141\,\AA\ line is still strong enough in most stars to be 
severely affected by microturbulence (Table\,\ref{Tabundchange})
and that e.g. Jehin et al. (1999) have picked systematically lower
microturbulence values than our choice of 1.5\,km/s. 
In fact, a lowering of our microturbulence parameters by 0.35\,km/s would
obliterate the mean difference relative to the just mentioned analyses.

Two stars appear Ba rich, the first one is the bright giant \object{HD\,196944}
with [Ba/Fe]\,$=+1.42$.
Za\u{c}s et al. (1998) found the star to be rich in $s$ elements and carbon,
with [Ba/Fe]\,$=+1.56$, while Aoki et al. (2002) derived [Ba/Fe]\,$=+1.10$,
and a very high lead abundance.
Van Eck et al. (2003) also find very high $s$-element abundances for this star.
Our second Ba rich star is \object{HD\,17072} with [Ba/Fe]\,$=+0.85$.
It was spectroscopically investigated by Carney et al. (1998), who in sharp
contrast found [Ba/Fe]\,$=-0.39$ using three Ba\,{\sc ii} lines.
They derived a 50\,K lower effective temperature, 0.6\,dex lower $\log g$,
a 0.6\,km/s higher microturbulent velocity, and a 0.2\,dex
lower iron abundance for the star.
Their measured equivalent width for the 6141.7\,\AA\ line of 147.6\,m\AA\,
is 12\,m\AA\ smaller than our measurement, and using only this line they
would have obtained [Ba/Fe]\,$=-0.61$, i.e., 1.46\,dex below our result.
Studying this extreme disagreement, we find that 0.24\,dex may be due to the
different microturbulent velocities, and 0.05\,dex due to the different values
of $\log g$, 0.02 and 0.03 from differing metallicities and temperatures, and
0.17\,dex is due to different $\log gf$ values
while the different equivalent widths only account for 0.16\,dex.
We can not trace the remaining difference.

\section{Discussion}
\subsection{Trends in relative abundance}
\begin{figure}
\centering
\resizebox{\hsize}{!}{\includegraphics*{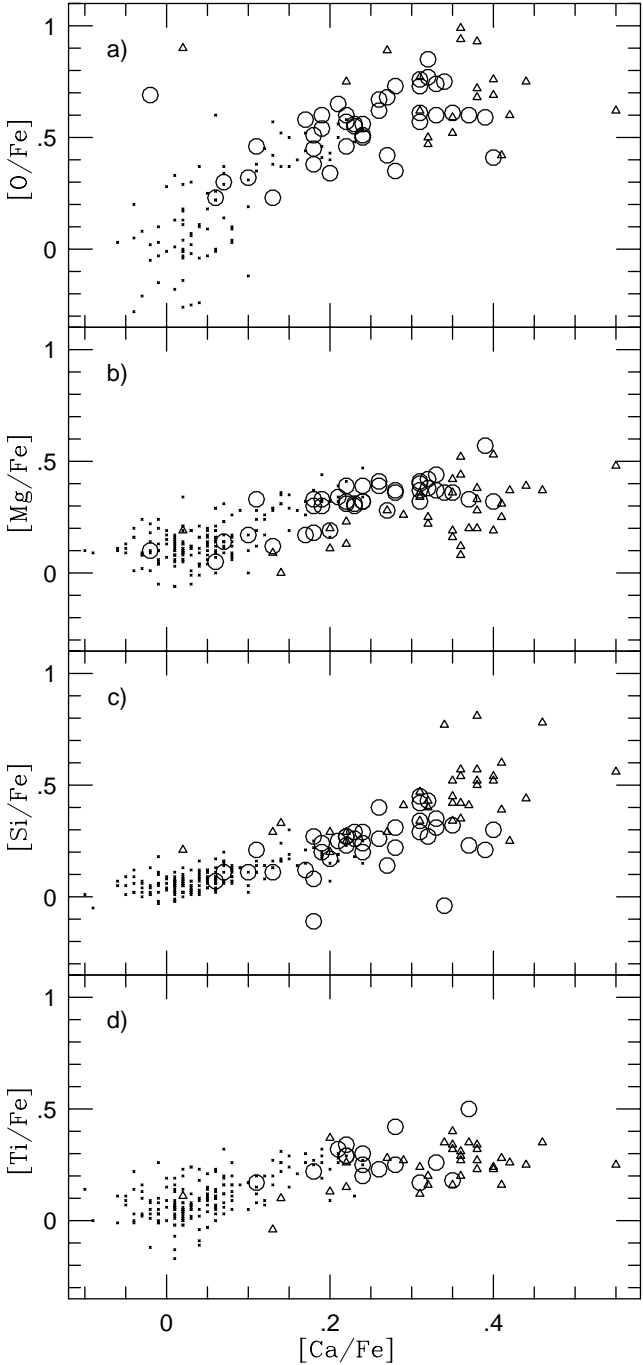}}
\caption[]{\label{FalphaCa}
Relative abundance relations of programme stars (circles, this paper),
disk stars (dots, Edvardsson et al. 1993), and extreme Pop\,II stars from
Cayrel et al. (2004) (triangles).
(Their [Fe/H]\,$=-4.0$ star CS\,22949-037 is not seen in panels a and b,
since it falls far above the other objects).
}
\end{figure}

The resulting NLTE abundances of oxygen relative to iron, as derived from the
O\,{\sc i} IR triplet lines, may systematically increase in [O/Fe] with
decreasing metallicity (Fig.\,\ref{FseveralprojectsOFe-FeH}) for Halo stars.
This tendency is, however, not significant.
Similar tendencies may possibly
also be traced for the $\alpha$ elements Mg (cf. also the discussion of 
NLTE effects in Sect.\,\ref{MgSiCa}) and Ca in Fig.\,\ref{Fresults}.  
We have explored whether this co-variation in abundances of the $\alpha$
elements may be observed if the metallicity [Fe/H] is left out explicitly.
In Fig.\,\ref{FalphaCa} we thus plot the abundances of O, Mg, Si and Ti relative
to Fe directly vs. [Ca/Fe] for our stars (circles), as well as for the disk
stars in Edvardsson et al. (1993) (dots), and find a relatively good agreement
in slopes between the two samples of stars for the different $\alpha$ elements.
We also note that we find similar slopes if we restrict our sample with
[Fe/H]\,$< -1.1$.
We have also extended the overall metallicity range in the figure by adding
the 35 extreme Pop\,II stars of Cayrel et al. (2004) (triangles)
of which all but two have metallicities [Fe/H] ranging from $-2.5$ to $-4.4$.
Their results are by necessity based on a very different set of lines
which are much too strong to be used in our investigation,
or by Edvardsson et al.
This may cause systematic differences in the results, especially the apparent
offsets of the triangles for Mg and Ti relative to Ca in panels b and d.

We have studied the correlations for the different data sets by calculating
the slopes of least-squares bisector fits (Babu \& Feigelson 1992).
These give the bisectors of two lines calculated as e.g.
[O/Fe]\,$=a+b$\,[Ca/Fe] and [Ca/Fe]$\,=c+d\,$[O/Fe].
The slopes $b$ and $1/d$ are often quite different, and the uncertainties in
the resulting bisector slopes are relatively large.
For [O/Fe] vs. [Ca/H] the slopes are $+3.0$, $+1.5$, $-1.3$, respectively,
for the data of Edvardsson et al., us and Cayrel et al.
For the combined data the slope is $+2.1$.
It would be interesting to test whether this apparent systematically changing
trend with the metallicity of the sample can be confirmed with independent data.

For the three other panels in Fig.\,\ref{FseveralprojectsOFe-FeH} the bisector
slopes are close to one, and show no obvious systematic differences
between the different samples.
This probably just indicates that the real scatters in Mg, Si, Ca and Ti vs. Fe
have indistinguishable properties within the present limits of measurement
accuracy.

Decauwer et al. (2005, Fig.\,6) have recently independently demonstrated
similar correlations among the different $\alpha$ elements for a smaller
sample of Pop\,II stars.

In trying to interpret the tendencies found we note that the core-collapse
supernova model yields, as calculated by
Woosley \& Weaver (1995), Umeda \& Nomoto (2002) and Chieffi \& Limongi (2004),
show systematically larger amounts of the lighter $\alpha$ elements relative to
the heavier ones as well as iron, the higher the SN initial mass.
That is, an Initial Mass Function varying with metallicity or time in the halo
such that the relative number of very massive stars was larger in the population
which enriched the early Halo, and with a gradual shift as the evolution
progressed,
could lead to the correlations observed.
Under the assumption that the $\alpha$ elements were made in
core-collapse supernovae we have made simple integrations of the yields of
Woosley \& Weaver (1995), Umeda \& Nomoto (2002) and Chieffi \& Limongi (2004)
for different values of a single-parameter IMF exponent.
These experiments show that in order for variations in the IMF for massive stars
to be able to explain the variations in Fig.\,\ref{FalphaCa}a, b and c, the
exponent has to decrease by two units, e.g., from $-1.5$ to $-3.5$ for the halo
stars.

For the disk, one could, as in Edvardsson et al. (1993) speculate that significant
fractions of Ca and Si were contributed by SNe of Type\,Ia or some other type of
objects with relative long life times, in addition to what is provided by the
short-lived stars giving rise to core collapse SNe.
This could also be an explanation for the different slopes in our plots.
It is possible that the characteristic time scale for SNe\,Ia to form is
relatively short as compared with the formation time of the Halo stars, in
particular if a fair fraction of them come from accreted dwarf
galaxies with slow star-formation rate (cf. Nissen \& Schuster 1997).
An important factor preventing SNe\,Ia to form may instead be low metallicity
(Kobayashi et al. 1998).
Alternative explanations for the slopes in Fig.\,\ref{FalphaCa} could
be departures from LTE, different for different $\alpha$ elements and varying
with [Fe/H]. This possibility should be explored systematically by detailed
NLTE calculations.

We have also investigated if our data show correlations similar to the
interesting relation between the abundances of Y, Ti and Fe found for Pop\,II
stars by Jehin et al. (1999).
Missing data for Y and with limited data for Ti, we looked for relations between
Ba, Ca and Fe abundances.
A tendency for [Ba/Fe] to increase with [Ca/Fe] for stars around
[Fe/H]\,$\sim -1.0$ was found, but in view of the uncertainties in the Ba
abundances we do not ascribe this tendency any clear significance. 

\subsection{Correlations between deviations from mean trends}
\label{Scorrelations}
\begin{figure}
\centering
\resizebox{4.5cm}{!}{\includegraphics*{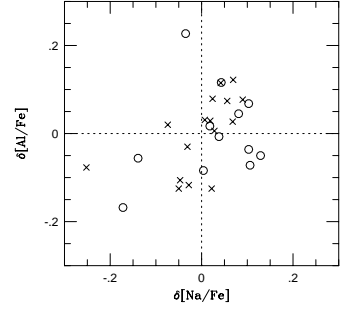}}
\caption[]{\label{FdAldNa}
Scatter correlation plot of Al and Na.
$\delta$[X/Fe] is the deviation in [X/Fe] from a typical [X/Fe] at that [Fe/H].
The circles and crosses represent, respectively, stars with [Fe/H] lower and
higher than $-1.0$.
}
\end{figure}

\begin{figure}
\centering
\resizebox{\hsize}{!}{\includegraphics*{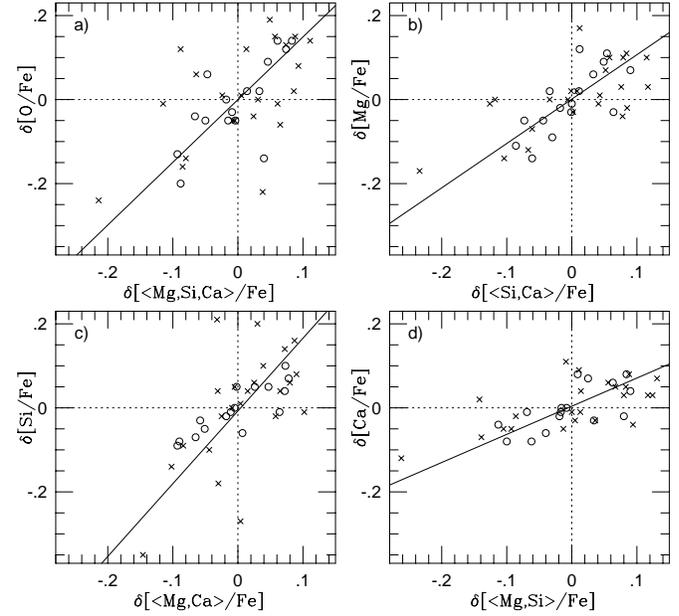}}
\caption[]{\label{Fdeltadelta}
Correlations among the scatters found in Fig.\,\ref{Fresults} for different
$\alpha$ elements X.
$\delta$[X/Fe] is the deviation in [X/Fe] from a typical [X/Fe] at that [Fe/H].
The circles and crosses represent, respectively, stars with [Fe/H] lower and
higher than $-1.0$.
Each line bisects the angle between two linear least-squares fits
$y=a+bx$ and $x=c+dy$ (Babu \& Feigelson 1992).
}
\end{figure}
We have computed broken mean relations similar to that in
Fig.\,\ref{FseveralprojectsOFe-FeH} to define the trends in Fig.\,\ref{Fresults}
and measured deviations from these in the vertical direction.
These deviations are denoted by $\delta$[X/Fe] for an element X.

An interesting first question is whether the scatter seen in the [Na/Fe] diagram 
(Fig.\,\ref{Fresults}b) is significant (so-called cosmic scatter), or whether it
may be explained as a result of errors in the analysis.
The scatter seems to increase for [Fe/H]\,$<-1.0$, but the Na lines
also get gradually weaker and may be more difficult to measure. 
We do not find any clear tendency for the deviation $\delta$[Na/Fe]
to correlate with
any other abundance deviation, but possibly with $\delta$[Al/Fe], at least in
the sense that the stars that seem most under-abundant in Na (for their [Fe/H])
are also under-abundant in Al, Fig.\,\ref{FdAldNa}.  This could 
contain information on synthesis; Na and Al are both odd nuclei and the
result of more complex nucleosynthesis in supernovae than the $\alpha$ elements.
We do not find any tendency for $\delta$[Na/Fe] to anti-correlate with
$\delta$[O/Fe], a tendency that might be present if the anomalies found for
some globular-cluster stars (e.g., Kraft et al. 1997)
were primordial and also represented among field halo stars.
We also do not find any clear tendency for $\delta$[Na/Fe] to vary
systematically with the kinematical characteristics of the stars.

An interesting fact is that different $\delta$[X/Fe] correlate for the different
elements X\,$=$\,O, Mg, Si, Ca and Ti. 
However, the slopes are not identical (Fig.\,\ref{Fdeltadelta}).
Similar effects show up for halo stars and disk stars, as is seen in
Fig.\,\ref{Fdeltadelta} where the stars are denoted differently, depending
on whether [Fe/H] is greater or smaller than $-1.0$.
It is important to study whether these correlations do not just reflect the
results of correlated more or less spurious errors. We first discuss the 
existence of the correlations as such, and then turn to the origin of the
different slopes.

One way of explaining the correlations would be to
advocate that mean errors in [Fe/H] of typically 0.10 would lead to slopes 
on the order of unity in Fig.\,\ref{Fdeltadelta}. However, most explanations 
for such errors would lead to correlated errors also in e.g. [Mg/H], so that the 
resulting
error in [Mg/Fe] would be reduced. E.g., spurious effective temperature errors 
of $\pm 150$\,K would lead to errors in [Fe/H] of about 0.10 but
less than half of that in [Mg/Fe] (Table\,\ref{Tabundcomp}).
Thus, with this approach errors considerably greater than 0.10\,dex in [Fe/H]
seem to be needed for explaining the correlations as such. 
Another more physical explanation for the scatter along lines with unity slope
in Fig.\,\ref{Fdeltadelta} would be to claim that there is a true cosmic scatter
in [X/Fe] for our Halo stars due to different contributions
of SNe Type\,Ia of iron. Thus, some stars would come from populations with such
a slow star-formation rate that these SNe have already contributed
significantly, resulting in comparatively low [$\alpha$/Fe] values,
while other stars (at the same [Fe/H]) come from regions still dominated by 
core-collapse SNe.
Although this explanation is not unrealistic, it would need further
confirmation, e.g. from dynamical arguments for a larger sample of stars.

Now we have to ask whether the different slopes for different elements X in
Fig.\,\ref{Fdeltadelta} could be the result of correlated errors.
Taking the spread along the line with unity slope as given, we wonder whether
the different errors in the
abundances for the different elements could spread the points from the line 
such that it leads to differences in slopes.
Qualitatively this may be so.
The slopes in Fig.\,\ref{Fdeltadelta} for O and Si is about 1.6
(panels a and c), while it is as low as 0.7 for Ca (panel d).
The fact that the slope decreases in progression from X = O and Si, Mg and Ca in 
Fig.\,\ref{Fdeltadelta} may be an indication pointing in this direction, since
we rate the quality of our relative abundances [X/Fe] as increasing in the same 
order. We have found from numerical experiments that considerable spurios errors
are needed, however, on the order of 0.2, 0.2, 0.15 and 0.10, respectively, in
[O/Fe], [Si/Fe], [Mg/Fe] and [Ca/Fe] to induce variations in the slopes as great
as observed.
Although not completely unrealistic, the results from our comparisons with other
studies above, as well as analysis of intrinsic errors, suggest 
only half as great abundance errors.
Thus there may be a more physical explanation for the correlations of 
Fig.\,\ref{Fdeltadelta}.

\begin{figure}
\centering
\resizebox{\hsize}{!}{\includegraphics*{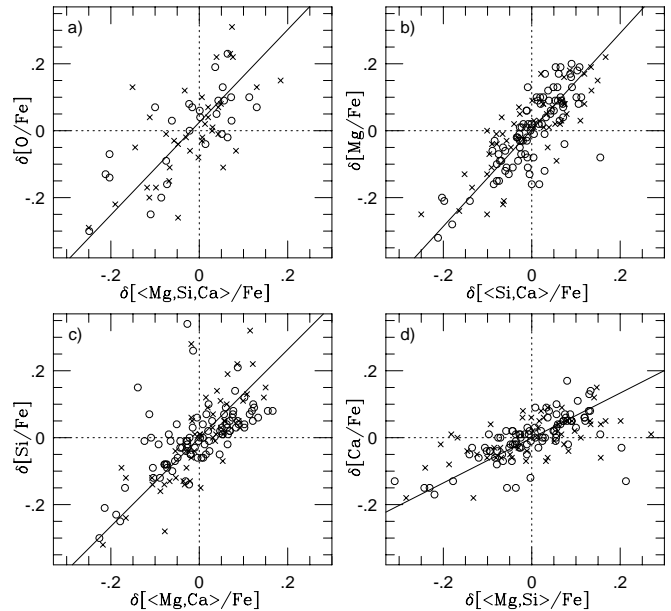}}
\caption[]{\label{FdeltaGratton}
As Fig.\,\ref{Fdeltadelta} but for the data of Gratton et al. (2003a).
}
\end{figure}

\begin{figure}
\centering
\resizebox{\hsize}{!}{\includegraphics*{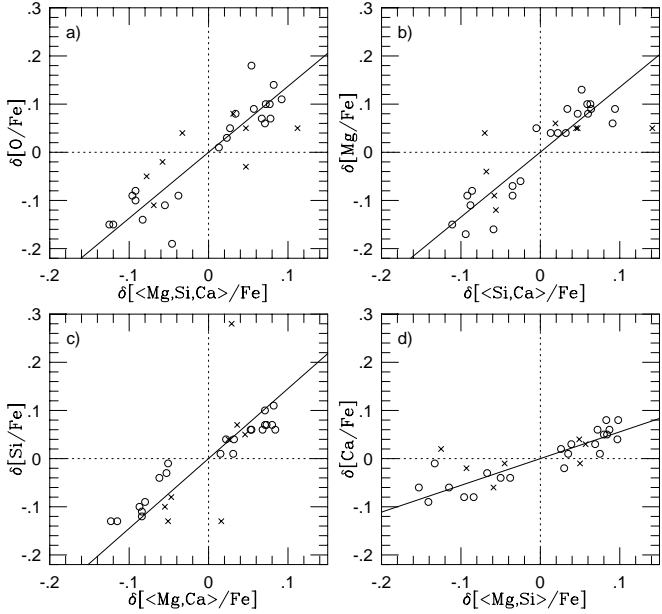}}
\caption[]{\label{FdeltaNissen}
As Fig.\,\ref{Fdeltadelta} but for the data of Nissen \& Schuster (1997).
}
\end{figure}

It is natural to explore whether the tendencies of Fig.\,\ref{Fdeltadelta} also
show up in the results of other studies.
We have investigated this for the studies of Gratton et al. (2003a) and
Nissen \& Schuster (1997) which were found (Sect.\,\ref{Sdiscussion}) to be
fairly consistent with ours and which contain
enough of determinations of O, Mg, Si and Ca abundances.
We have performed the analyses for these two studies independently, plotting
the stars in the [X/Fe] vs. [Fe/H] diagrams with X= Mg, Si and Ca, fitting
lines broken at [Fe/H]\,$= -1.0$ to the ``tendencies'', reading off
$\delta$\,[X/Fe] and plotting the diagrams corresponding to
Fig.\,\ref{Fdeltadelta}.
All stars with data were used, not just those overlapping with our sample -- in
total, there were 150 stars from Gratton et al. (2003a) and 30 from
Nissen \& Schuster (1997).
The resulting diagrams are displayed in Figs.\,\ref{FdeltaGratton} and
\ref{FdeltaNissen}.

It is seen that very similar correlations to those of ours do result.
When the full sample of measurements are used, we find the mean slopes in these
diagrams as given in Table\,\ref{Tslopes}.
\begin{table}
\caption[]{\label{Tslopes}
Slopes in correlation diagrams similar to Figs.\,\ref{Fdeltadelta} to
\ref{FdeltaNissen}, for the added sample of measurements by us,
Gratton et al. (2003a) and Nissen \& Schuster (1997).
Data for the subsamples with [Fe/H]\,$> -1.0$ and $\le -1.0$
are given separately, as well as for the full sample.
}
\begin{flushleft} 
\begin{tabular*}{8.8cm}{lccc}
Diagram & \multicolumn{3}{c}{[Fe/H]} \\
          \cline{2-4} \\
        & $> -1.0$ & $\le -1.0$ & All \\
\hline
 $\delta$[O/Fe]/$\delta$[$<$Mg,Si,Ca$>$/Fe] & 1.77 & 1.79 & 1.84 \\
$\delta$[Mg/Fe]/$\delta$[$<$Si,Ca$>$/Fe]    & 1.26 & 1.66 & 1.48 \\
$\delta$[Si/Fe]/$\delta$[$<$Mg,Ca$>$/Fe]    & 1.59 & 1.04 & 1.25 \\
$\delta$[Ca/Fe]/$\delta$[$<$Mg,Si$>$/Fe]    & 0.59 & 0.67 & 0.64 \\
\hline
\end{tabular*}
\end{flushleft}
\end{table}
   
It is noteworthy that the slopes in the [O,Mg,Si,Ca/Fe] vs. [Fe/H] diagrams
of Edvardsson et al. (1993) (cf. also Fig.\,\ref{Fresults} above) for disk
stars show a rather similar tendency to the slopes in Figs.\,\ref{Fdeltadelta},
\ref{FdeltaGratton} and \ref{FdeltaNissen} and Table\,\ref{Tslopes}, with a
slope for O being about 1.4 times that of
the $\alpha$ elements, while the slope for Ca is about 0.70 of the mean slope.
(As mentioned above, in Edvardsson et al. this slope difference was tentatively 
interpreted as the result of variations in the fractions of elements
contributed by different types of supernovae.)
The similarity between the slopes in the panels of Fig.\,\ref{Fdeltadelta} and
those in Edvardsson et al. becomes natural if our halo stars were contributed
from different environments where star formation has proceeded at different
rates with different relative contributions from SNe\,Ia.

Another possibility would be that we see the result from spurious variations in
the IMF at different places in the early Galactic Halo.
There is increasing evidence that the IMF varies significantly for star
formation occurring in different environments. Thus, a uniform single-parameter
IMF, such as the Salpeter IMF, may merely reflect a statistical mean of stars
formed in many different regions, while locally the IMF may vary
significantly, see, e.g., Elmegreen (2004) and references therein.
A type of variable-IMF effect suggested to explain relative-abundance 
differences similar to ours is that of
Jehin et al. (1999) and Decauwer et al. (2005).
In the latter paper it is proposed that even with a universal single-parameter
IMF, a low-mass star-forming region does not form enough stars to make even a
single very massive star, which could later expel very much oxygen.
The subsequent generation of stars, born in the same cloud after the first epoch
of core-collapse supernovae, would then be less oxygen-enriched than stars
formed in a more massive cloud where also very massive stars could contribute
extra oxygen.
This scenario might be able to explain the individual star-to-star scatter seen
in Fig.\,\ref{Fdeltadelta}. 
It is, however, at odds with the observations for more metal-poor halo stars of
Arnone et al. (2004) of very small scatter in relative abundances.

There is also a possibility that the slope differences for different elements,
in Fig.\,\ref{Fdeltadelta} and/or those discussed by Edvardsson et al. (see
above), merely reflect systematic errors, e.g. varying with metallicity in the
latter case. 
This could then reflect effects of departures from LTE or possibly convection 
inhomogeneities.
The effects shown in Fig.\,\ref{Fdeltadelta} are,
however, difficult to explain as being due to errors in effective temperatures
or other model parameters as may be found from Table\,\ref{Tabundchange}.
We conclude that another stellar parameter, e.g. related to rotation or magnetic
field, may be needed to explain the correlations as results of spurious errors
instead of real abundance differences.

\subsection{Abundances and kinematics}
We divide our metal-poor stars ([Fe/H]\,$<-1.0$) into two subsamples, roughly
corresponding to the class ``accretion component'' of Gratton et al. (2003b),
here $V_{\rm LSR} < -200$\,km/s (9 stars), and their ``dissipative collapse
component'', here $V_{\rm LSR} > -200$\,km/s (17 stars).
Gratton et al. suggest that the accretion component stars have a lower mean and
a larger spread in [$\alpha$/Fe] than those of the dissipative collapse
component.
We find the following differences in mean [X/Fe], $\Delta$$<$[X/Fe]$>$, and the
ratios, $R_{\sigma({\rm [X/Fe]})}$, of the standard deviation of the scatter
between the accretion and dissipative components:
$\Delta$$<$[Mg/Fe]$>\,=-0.03 \pm 0.06$, $R_{\sigma({\rm [Mg/Fe]})}=1.7 \pm 0.5$,
$\Delta$$<$[Si/Fe]$>\,=-0.07 \pm 0.08$, $R_{\sigma({\rm [Si/Fe]})}=1.4 \pm 0.5$
and
$\Delta$$<$[Ca/Fe]$>\,=+0.01 \pm 0.07$, $R_{\sigma({\rm [Ca/Fe]})}=3.2 \pm 1.1$.
These $R$ values are statistically significant and consistent with Fig.\,6 of
Gratton et al. (2003b), but even if our three $\Delta$$<$[X/Fe]$>$ values are
considered together, they are not significantly different from zero.
For oxygen, which is presumably formed in the same environments as the $\alpha$
elements, the corresponding figures are
$\Delta$$<$[O/Fe]$>\,=-0.05 \pm 0.06$, $R_{\sigma({\rm [O/Fe]})}=1.1 \pm 0.35$.

\begin{figure}
\centering
\resizebox{\hsize}{!}{\includegraphics*{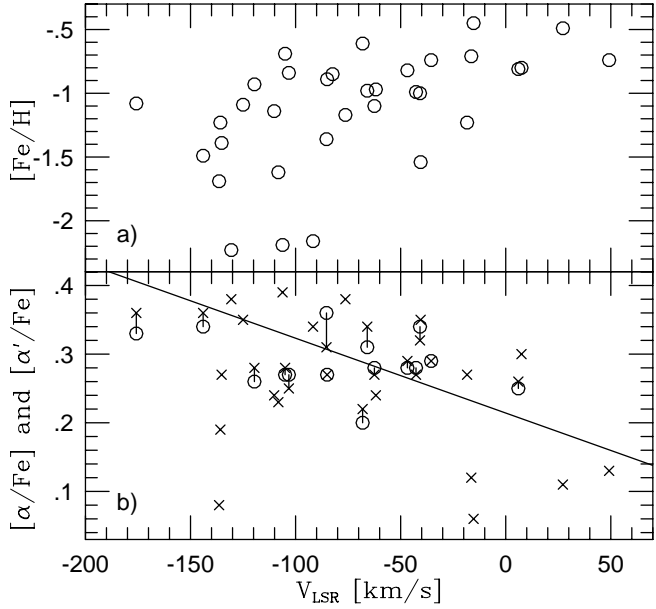}}
\caption[]{\label{FFe_alpha_V}
Correlations of iron abundances and [$\alpha$/Fe] with $V_{\rm LSR}$ for stars
of the ``dissipative collapse component'' of Gratton et al. (2003b).
In panel b the line shows Eq.\,3 of Gratton et al., the circles denote
[$\alpha$/Fe]\,$=\frac{1}{4}({\rm [Mg/Fe]}+{\rm [Si/Fe]}+
{\rm [Ca/Fe]}+{\rm [Ti/Fe]})$ and the crosses
[$\alpha$'/Fe]\,$=\frac{1}{3}({\rm [Mg/Fe]}+{\rm [Si/Fe]}+{\rm [Ca/Fe]})$.
          }
\end{figure}

Finally, in Fig.\,\ref{FFe_alpha_V} we may probably trace the tendencies, found
by Gratton et al. (2003b), that stars of the dissipative component show
increasing values of [Fe/H] and decreasing [$\alpha$/Fe]
with increasing rotational velocity $V$.

\section{Conclusions}
We have determined chemical abundances for 43 metal-poor solar-type stars, 
most of them dwarfs or subgiants from the Galactic Halo and old Disk.
The abundances have been determined for 13 elements, but for 4 elements only 
one spectral line could be used. 

The analysis has been strictly differential relative to the Sun, based on 1D 
LTE model atmospheres both for the Sun and for the stars. Since the Sun is
more metal-rich and thus different from all programme stars,
systematic errors on the order of 0.15\,dex might occur.
Another problem may be the effective-temperature scale for which an estimated
error of about 100\,K may cause systematic errors on the order of 0.07\,dex.
Other important sources of error, which have not been explored here,
are effects of convective inhomogeneities and departures from LTE.
The latter were, however, considered in deriving the oxygen abundances from the
O\,{\sc i} IR triplet lines.

No overionisation of Fe, contrary to what was suggested by
Th\'evenin \& Idiart (1999), was found from our single Fe\,{\sc ii} line, which
agrees with the finding of Gratton et al. (2003a).
Comparisons with other similar studies show agreement for stars in common,
with mean differences in abundances of typically significantly less than
0.1\,dex.
For the elements only represented by one line, the differences may be larger.

We have found relations of [X/Fe] vs. [Fe/H], with X = O, Na, Mg, Al, Si, Ca,
Ti, and Ni, similar to those that have been found by others.
Thus, for X = O we found a slight increase with decreasing [Fe/H], though much
smaller than previously found from the ultraviolet OH bands.
Na behaves somewhat differently compared to Al, for which
there is an increase, similar to that of Mg, Si, Ca and Ti.
There are some indications
that the relative abundance ratios of the $\alpha$ elements vary among the halo
stars, and that some $\alpha$ abundances relative to Fe may increase
with decreasing metallicity.  
Ni behaves like Fe, as also Sc and Cr seem to do. 

We have traced a scatter around the mean relations in the [X/Fe] vs. [Fe/H]
diagrams that may be partly real. There seem to be correlations between the
deviations in [X/Fe] from the mean trends for different $\alpha$ elements X, and
the coefficients in these correlations are different. This might be due to 
larger spurious errors than anticipated and understood, but may also have a more
physical explanation. The slope differences are similar to those in the
[X/Fe] vs. [Fe/H] diagrams for disk stars. This could result if the halo
stars were formed in regions with different (and considerable) characteristic
star-formation times.
Alternatives would be a widely varying IMF among the halo-star forming clouds
or a bias against the formation of very massive stars in a fraction of the
clouds.
We have found support for the finding by Gratton et al. (2003b) 
that stars that do not participate in the rotation of the galactic disk 
show a lower mean and larger spread in [$\alpha$/Fe], than stars participating
in the general rotation which also show some correlation between [$\alpha$/Fe]
and rotation speed. 

We also have identified a number of stars with departing or peculiar
abundances, including two Ba stars -- one of them known before, and two stars
that seem to have low [Ni/Fe] abundances. 

The present study is not optimal -- a considerably larger sample,
with spectra at higher S/N and wider wavelength coverage can be reached, even
for more metal-poor and thus fainter stars. The analysis can be improved,
by using hydrogen line profiles to set the temperature scale, and by
using contemporary 3D hydrodynamical models and detailed NLTE modelling.
In spite of the existence of fairly consistent results for many stars covered by
recent studies, a new even larger homogeneous abundance survey is important.
In view of the unsolved problems concerning the origin of the Halo and the 
early nucleosynthesis, as well as the character of the ``Halo-Disk transition"
and the unknown origin of the probably physical scatter along the mean abundance 
trends, such a study is in fact needed.
When designing such a study, one should 
put emphasis on the selection criteria applied -- until now, in all the studies
of general Halo stars known to us the kinematic criteria have been introduced
in a way which is more or less ad-hoc.

\begin{acknowledgements}
We thank Dan Kiselman (Stockholm) for kindly providing an oxygen
model atom for calculating departures from LTE for the oxygen triplet lines,
Harri Lindgren (ESO) for making available unpublished photometric
and radial-velocity data, Chen Yuqin (Beijing) for stellar gravity calculations.
Accurate radial velocities from CORAVEL data were kindly provided in advance of
publication by Birgitta Nordstr\"om (Copenhagen and Lund).
Paul Barklem (Uppsala), Sofia Feltzing (Lund) and Andreas Korn (Uppsala) are
thanked for valuable critical
comments on an earlier version of the manuscript, and Torgny Karlsson
(Copenhagen) for valuable discussions.
The anonymous referee is thanked for many contributions and improvements.

We have made use of data from the Hipparcos mission, the NASA ADS and the VALD
databases.  This research has also made use of the SIMBAD database, operated at
CDS, Strasbourg, France.

This project has been supported by the Swedish Natural Science Research Council
(NFR, prior to A.D. 2001) and the Swedish Research Council (VR).

\end{acknowledgements}

\end{document}